\documentclass[preprint2]{aastex}
\usepackage{color}
\definecolor{Dred}{rgb}{0.312,0.070,0.070}
\definecolor{Dblue}{rgb}{0.070,0.070,0.312}
\definecolor{Dgreen}{rgb}{0.070,0.312,0.070}
\definecolor{Db}{rgb}    {0.050,0.0,0.320}

\newcounter{note}
\let\oldmarginpar\marginpar
\renewcommand\marginpar[1]{\-\oldmarginpar[\raggedleft\footnotesize #1]{\raggedright\footnotesize #1}}

\shorttitle{iMOGABA: S5 0716+714}
\shortauthors{Lee et al.}

\begin{document}

\title{Interferometric Monitoring of Gamma-ray Bright AGNs:
S5~0716+714}

\author{
Jee Won Lee\altaffilmark{1,2},          %
Sang-Sung Lee\altaffilmark{1,3},        %
Jeffrey A. Hodgson\altaffilmark{1},     %
Dae-Won Kim\altaffilmark{4},
Juan-Carlos Algaba\altaffilmark{1},     %
Sincheol Kang\altaffilmark{1,3},        %
Jiman Kang\altaffilmark{1}, and
Sungsoo S. Kim\altaffilmark{2}            %
}

\altaffiltext{1}{
Korea Astronomy and Space Science Institute, 
776 Daedeok-daero, Yuseong-gu, Daejeon 34055,
Republic of Korea; Correspondence to Sang-Sung Lee at sslee@kasi.re.kr}

\altaffiltext{2}{
Department of Astronomy and Space Science,
Kyung Hee University, 1732, Deogyeong-daero,
Giheung-gu, Yongin-si, Gyeonggi-do 17104,
Republic of Korea
}

\altaffiltext{3}{
University of Science and Technology, 
217 Gajeong-ro, Yuseong-gu, Daejeon 34113, 
Republic of Korea}

\altaffiltext{4}{Department of Physics and Astronomy,
Seoul National University, 
1 Gwanak-ro, Gwanak-gu, Seoul 08826,
Republic of Korea}

\begin{abstract}
We present the results of very long baseline interferometry (VLBI) observations of 
gamma-ray bright blazar S5 0716+714
using the Korean VLBI Network (KVN)
at the 22, 43, 86, and 129~GHz bands,
as part of the
Interferometric Monitoring of Gamma-ray Bright AGNs (iMOGABA) KVN key science program.
Observations were conducted in 29 sessions from January 16, 2013 to March 1, 2016, with the source being detected and imaged at all available frequencies.
In all epochs, the source was compact on the milliarcsecond (mas) scale, 
yielding a compact VLBI core
dominating the synchrotron emission
on these scales.
Based on the multi-wavelength data between 15~GHz (Owens Valley Radio Observatory) and 230~GHz (Submillimeter Array), we found that
the source shows multiple prominent enhancements of the flux density
at the centimeter (cm) and millimeter (mm) wavelengths,
with mm enhancements leading cm enhancements by -16$\pm$8 days.
The turnover frequency was found to vary between 21 to 69\,GHz during our observations. 
By assuming a synchrotron self-absorption model for the relativistic jet emission in S5 0716+714,
we found the magnetic field strength
in the mas emission region to be 
$\le5$~mG during the observing period, yielding 
a weighted mean of $1.0\pm0.6$~mG for higher turnover frequencies (e.g., $>$45~GHz).

\end{abstract}


\section{Introduction}\label{introduction}

A radio-loud active galactic nucleus (AGN) is a very compact region in a galaxy 
 with a super-massive black hole
(SMBH, $M_{\rm BH}\approx 10^{6}-10^{9} M_{\odot}$) in its central part, 
emitting at extremely high luminosity with a relativistic jet (Radio-quiet AGNs also sometimes have a jet)~\citep[see e.g.,][]{Krolik+99}.
Blazars are a subclass of AGN with a relativistic jet directed toward the observer (i.e., small viewing angle). Some of the major physical properties of blazars across the entire electromagnetic spectra from radio to gamma-rays include flux density variability on time scales of hours to years~\citep[see e.g.,][for review]{Boettcher+12}.

The observed variability suggests a possible connection to different regions of the jet e.g., rapid variations being connected to the innermost regions of relativistic jets, and slow variations being related to downstream jet emission (on parsec scales). 
High-resolution multi-frequency very long baseline interferometry (VLBI)
observations enable us to investigate synchrotron radiation. 
Radio emission is thought to be due to synchrotron emission from a relativistic jet 
on the milliarcsecond (mas) scale, yielding a plethora of observables such as flux density, source structure, and polarization properties, which can help us to study the physical conditions in the jet in the vicinity of the SMBH.

S5 0716+714 is one of the most active blazars known, with a flat spectrum, located at z $\sim$ 0.127 \citep{Stadnik+14}. The remarkable properties of this source are its extreme flux density variability, such as broadband flaring on a 
time scale of months~\citep{Raiteri+03, Nesci+05, Rani+13} and its intraday variability (IDV), possibly from an extrinsic origin for the cm-wavelength IDV~\citep{Wagner+96, Liu+12} and from an intrinsic origin for mm-wavelength IDV~\citep{Kraus+03, Leejw+16}.
\cite{Bhatta+15} detected five-hour-long microflares in flux density and polarization in optical band with the Whole Earth Blazar Telescope (WEBT) during the iMOGABA observing campaign.
This was interpreted by propagation and compression of a small-scale but strong shock within the jet.
\cite{Marti-Vidal+16} computed core-shift between 14.4\,GHz and 43\,GHz by VLBI observations for S5 0716+714. 
They found that the core shift is roughly aligned with the jet direction because the core shift shall be aligned with the highest magnetic field and/or particle density gradient, which are given in the direction longitudinal to the jet.

In general, flares lasting approximately months are explained by the shock-in-jet model~\citep{Marscher+85}.
VLBI studies of the source on the mas scales show a core-dominated jet \citep{Bach+05,Rastorgueva+11}
and suggest a connection between the jet kinematics and the observed broadband flares~\citep{Rani+15}.
Based on mm-VLBI observations and gamma-ray monitoring with \textit{Fermi}/LAT (Large Area Telescope) of S5 0716+714,
it was reported that variation of the gamma-ray flux is correlated with position angle variations in the VLBI jet.
Additionally, they found variation in the mm-VLBI core flux density, indicating that the high-energy emission originates upstream of the mm-VLBI core~\citep{Rani+14}. In a broadband long-term flare study of S5 0716+714,
\cite{Rani+13} found that flares in the jet propagate from the high-energy emission region to the low-energy emission regions
with a time delay of $\sim$65 days due to opacity, yielding
the evolution of the spectra following the shock-in-jet model, 
the emission region size of a few $\mu$as,
and a magnetic field of a few mG.
However, these multi-frequency studies may have possible uncertainty in the derived spectral information for S5 0716+714, as the multi-frequency observations were conducted only quasi-simultaneously. 

In this paper, we report the results of simultaneous multi-frequency
VLBI monthly monitoring observations of S5 0716+714 over three years
using the Korean VLBI Network (KVN) at 22, 43, 86, and 129~GHz.
These observations enable us to investigate the spectral information
of the synchrotron emission regions on the mas scale.
In Section~\ref{obs_reduction}, we describe the VLBI observations with KVN
and the data reduction procedures used,
and in Section~\ref{data}, we summarize the multi-wavelength data collected for our study. We present the results of the observations and analysis in Section~\ref{results}, followed by the discussion in Section~\ref{discussion}.
Finally, a summary of the paper is presented in Section~\ref{summary}.

\section{Observations and Data Reduction\label{obs_reduction}}

\subsection{Observations\label{obs}}

VLBI observations of S5 0716+714 were performed as part of
the Interferometric MOnitoring of Gamma-ray Bright AGNs (iMOGABA)
project~\citep[][]{Lee+13,Lee+16}.
More than 30 radio-loud AGNs are monitored monthly and
simultaneously at 22, 43, 86, and 129~GHz,
with the first observation occurring on December 5, 2012 (MJD 56266),
using the Korean VLBI Network (KVN)~\citep{Lee+11,Lee+14}.
The KVN is a 500-km VLBI array consisting of three identical
21-m radio telescopes:
the KVN Yonsei (KY), the KVN Ulsan (KU), and the KVN Tamna (KT).
The typical spatial resolutions are 6, 3, 1.5, and 1 milliarcsecond (mas)
at 22, 43, 86, and 129~GHz, respectively.

The iMOGABA observations used for this paper were conducted
during a period of more than three years from January 16, 2013 (MJD 56308) to March 1, 2016 (MJD 57448) with a mean cadence of $\sim$30 days.
All sources were observed in every 24-hour session
with 1-8 five-min scans.
The observing frequencies are 21.700--21.764~GHz, 43.400--43.464~GHz,
86.800--86.864~GHz, and 129.300--129.364~GHz,
in left circular polarization (LCP),
yielding a total bandwidth of 256~MHz and a bandwidth of 64~MHz for each band (when observed at four frequency bands).
The starting frequencies of each band have integer-ratios
for the application of the frequency phase transfer (FPT) technique
for faint sources. The FPT technique was successfully demonstrated for iMOGABA observations by \cite{Algaba+15}. To trace the opacity of the atmosphere during each 24-hr observation, hourly sky tipping curve measurements were performed at each antenna every hour by measuring the system temperatures at the following eight elevations:
18.21$^\circ$, 20.17$^\circ$, 22.62$^\circ$, 25.77$^\circ$, 30.00$^\circ$,
36.03$^\circ$, 45.58$^\circ$, and 65.38$^\circ$.
Antenna pointing offsets were corrected for every scan through cross-scan observations of target sources.
During each scan, the received signals were digitized (i.e., 2-bit quantized) by digital samplers,
requantized, divided into 16 sub-bands (IFs) with a bandwidth of 16~MHz by a digital filter bank,
and then recorded by the Mark 5B system at a rate of 1024~Mbps.
The recorded data were correlated using the DiFX software correlator
in the Korea-Japan Correlator Center~\citep{Deller+07,Lee+15a}.

\subsection{Data analysis\label{dr}}

The output from the DiFX correlator is the spectrum
of the cross-correlation function with a resolution of 0.125~MHz
and an accumulation period of 1~s.
A post-correlation process was conducted using
the  National Radio Astronomy Observatory (NRAO) Astronomical Image Processing System (AIPS).
Post-correlation processing was done both manually
and automatically using the KVN Pipeline.
Data were reduced using standard AIPS tasks and with ParselTongue \citep{Kettenis+06} with the operation of the KVN Pipeline being described in detail by \citet{Hodgson+16}.
Sensitivity at high frequencies was improved by transferring phase solutions
from longer to shorter wavelengths (i.e., FPT) as described by
\cite{Rioja+11}, \cite{Rioja+14}, and \citet{Algaba+15}.

An antenna-based fringe fit was conducted using the AIPS task FRING.
The 5-min scans were split into solution intervals of 30~s.
Baseline-based fringe solutions (phase delays and delay rates)
were solved for individual IFs at all frequency bands.
The baseline-based fringe solutions were used to determine
the antenna-based fringe solutions.

Amplitude calibration was performed using 
the system temperatures measured during the observations.
The measured system temperatures were corrected
for atmospheric opacity as determined by sky tipping curve measurements conducted every hour during
the observations at each telescope.
Re-normalization of the fringe amplitudes was done to correct for amplitude distortion due to quantization,
and the quantization and re-quantization losses were
corrected~\citep{Lee+15a,Lee+15c}.
After the amplitude calibration and the re-quantization loss correction,
we expect that the uncertainty of the amplitude calibration is
within 5~\% at 22 and 43~GHz bands~\citep{Petrov+12,Lee+14,Lee+15c}
and 10-30~\% at 86 and 129~GHz bands~(Lee et al. in prep.).

\subsection{Imaging and model-fitting\label{imaging-modelfit}}
Although there are limitations in imaging using visibility data
obtained with only three antennas, i.e., only one closure phase
for each scan and no closure amplitude,
the hybrid mapping for the KVN observations is feasible for discussing the total flux spectral index of such a compact radio source.
CLEAN images were produced using the Caltech DIFMAP software.
The calibrated {\it uv}-data were averaged in terms of the frequency by chopping the band edges (10\% of the bandwidth),
where severe flux loss arises
due to the bandpass shape.
The data were then re-averaged in time over 30 seconds at 22, 43, and 86\,GHz and over 10 seconds at 129\,GHz, taking into account the coherence time at each frequency band, and expecting decoherence loss of up to 30\,\% at 86-129\,GHz bands~(Lee et al. in prep.).
Outliers in the amplitude of the averaged {\it uv}-data were flagged out.
For some epochs, the {\it uv}-data contained considerable antenna pointing offsets due to anomalous refraction,
which caused significant antenna-based flux loss.
A single circular Gaussian model was used to fit and phase-self-calibrate
the averaged {\it uv}-data.
The self-calibrated data were again fitted with a set of CLEAN point source
models and self-calibrated with respect to the CLEAN models.
The iteration  of CLEAN and self-calibration were run until the residual noise in the CLEAN image is comparable to the Gaussian random noise. 

\begin{figure*}[!t]
\epsscale{0.9}
\plotone{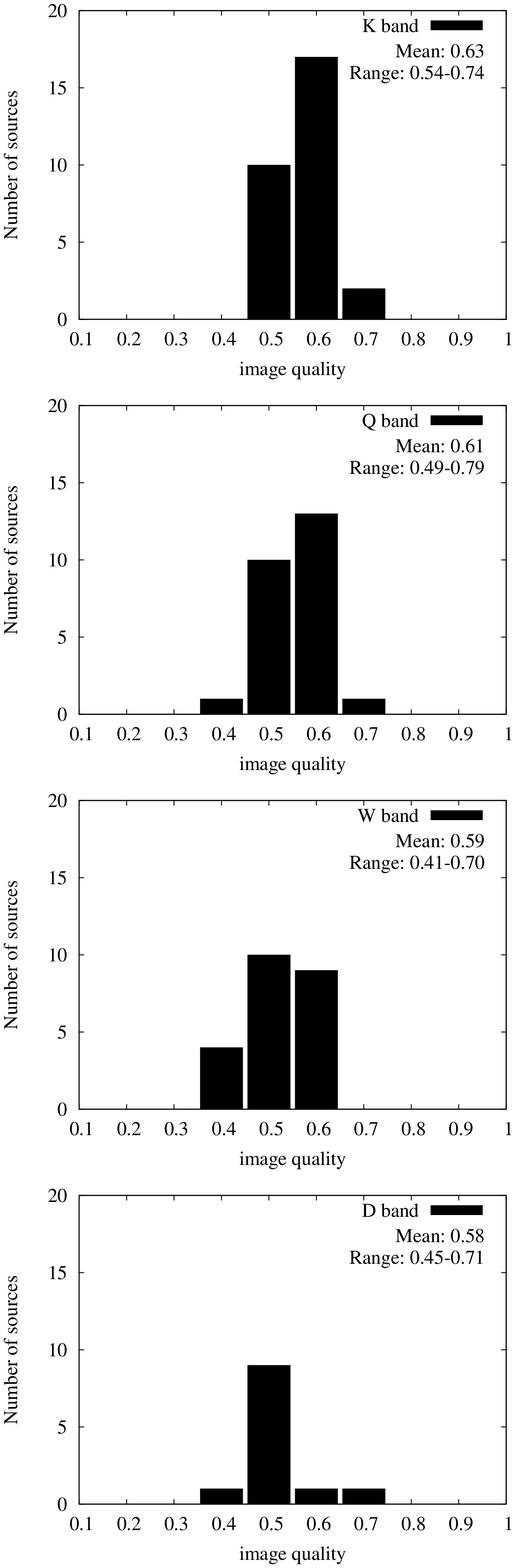}
\caption{The distribution of the image quality factor $\xi_{\rm r}$ (ratio of the image root-mean-square noise to its mathematical expectation).
The mean and range of the distribution are presented.
\label{fig-hist-q}}
\end{figure*}

The residual noise was investigated to evaluate the image quality, as
described in \cite{Lee+16}.
Under the assumption
that the residual noise of the ideal CLEAN image is random noise
and hence should have a Gaussian distribution,
the noise in the final CLEAN image could be evaluated using 
an image quality factor, i.e.,
the ratio of the image noise RMS to its mathematical expectation,
$\xi_{\rm r} = s_{\rm r} / s_{\rm {r,exp}}$,
where $|s_{\rm r}|$ is the maximum absolute flux density in the residual image
and $|s_{\rm r,exp}|$ is the expectation of $s_{\rm r}$. 
For residual random Gaussian noise with a zero mean,
the expectation of $s_{\rm r}$ is
$|s_{\rm {r,exp}}| = \sigma_{\rm r} { \left[ \sqrt{2} \ln{\left( \frac{N_{\rm pix}} {\sqrt{2\pi}\sigma_{\rm r}} \right) }\right]}^{1/2}$,
where $N_{\rm pix}$ is the total number of pixels in the image
and $\sigma_{\rm r}$ denotes the image noise rms.
When the residual noise is close to Gaussian noise, the image quality factor $ \xi_{\rm r} \rightarrow 1 $;
when the residual image still contains a source structure, 
$ \xi_{\rm r} > 1 $;
and when the final image has
a large number of degrees of freedom, $ \xi_{\rm r} < 1 $~\citep{Lobanov+06}.
We found that the values of $\xi_{\rm r} $ of the final CLEAN images
obtained in this paper
are in the range of
0.54-0.74 at 22~GHz,
0.49-0.79 at 43~GHz,
0.41-0.70 at 86~GHz,
and 0.45-0.71 at 129~GHz,
as shown in Figure~\ref{fig-hist-q}
and as summarized in Table~\ref{t2},
implying that the images adequately represent the structure 
observed.

Given that the target source S5 0716+714 is a circumpolar source,
the {\it uv}-distribution fill only a limited range, i.e.,
the lack of relative short baseline giving rise to a lack of data
for the innermost region in the {\it uv}-plane.
This may cause a missing-flux problem in the CLEANed total flux density: 
a larger-scale structure of the source may be resolved out
and hence the flux density for the larger-scale structure may miss.
However, for such a compact radio source on mas-scales to arcsecond-scales,
the missing-flux effect may not be significant.
We expect that for future studies,
single-dish (zero spacing) observations should be accompanied
for properly measuring the total flux density of the source
free from the missing-flux effect.

The visibility data of the final CLEAN images
were used for fitting with circular Gaussian models.
From the Gaussian model-fitting step, a set of fit parameters were obtained, as follows:
{\it$S_{\rm tot}$} (total flux density), {\it $S_{\rm peak}$} (peak flux density
), 
{\it $\sigma_{\rm rms}$} (post-fit rms), $d$ (size), 
$r$ (radial distance, for jet components), and
$\theta$ (position angle, measured for jet components, with respect to the location of the core component), as summarized in Table~\ref{t3}.

The uncertainties in Table~\ref{t3}
were obtained while taking into account the signal-to-noise ratio during the detection
of each component,
and following \cite{Lee+16}.
The minimum resolvable size was estimated for each component according to \cite{Lobanov+05}.
The minimum resolvable size was used to determine then upper limit of the component size 
when the fitted component size $d$ was smaller than the minimum resolvable size.
The upper limit of the size was also used to estimate
the lower limit of the brightness temperature $T_{\rm b}$, as established by 
$T_{\rm b} = \frac{2\ln{2}}{{\pi}{k}}\frac{S_{\rm tot}\lambda^2}{d^2}(1+z)$,
where $\lambda$ is the wavelength of the observation, $z$ is the redshift, 
and $k$ is the Boltzmann constant.

\subsection{Images and model-fit parameters}

We detected S5 0716+714 during 29 observing sessions
for an observing period of more than
three years, from January 16, 2013 to March 1, 2016, 
in at least one frequency band, yielding 89 VLBI images.
These were 29 images at 22~GHz, 25 images at 43~GHz, 23 images at 86~GHz,
and 11 images at 129~GHz.
These are the first simultaneous multi-frequency VLBI long term monitoring images,
yielding a very compact core-dominated structure on the mas scale.

\begin{figure*}[!t]
\epsscale{2}
\plotone{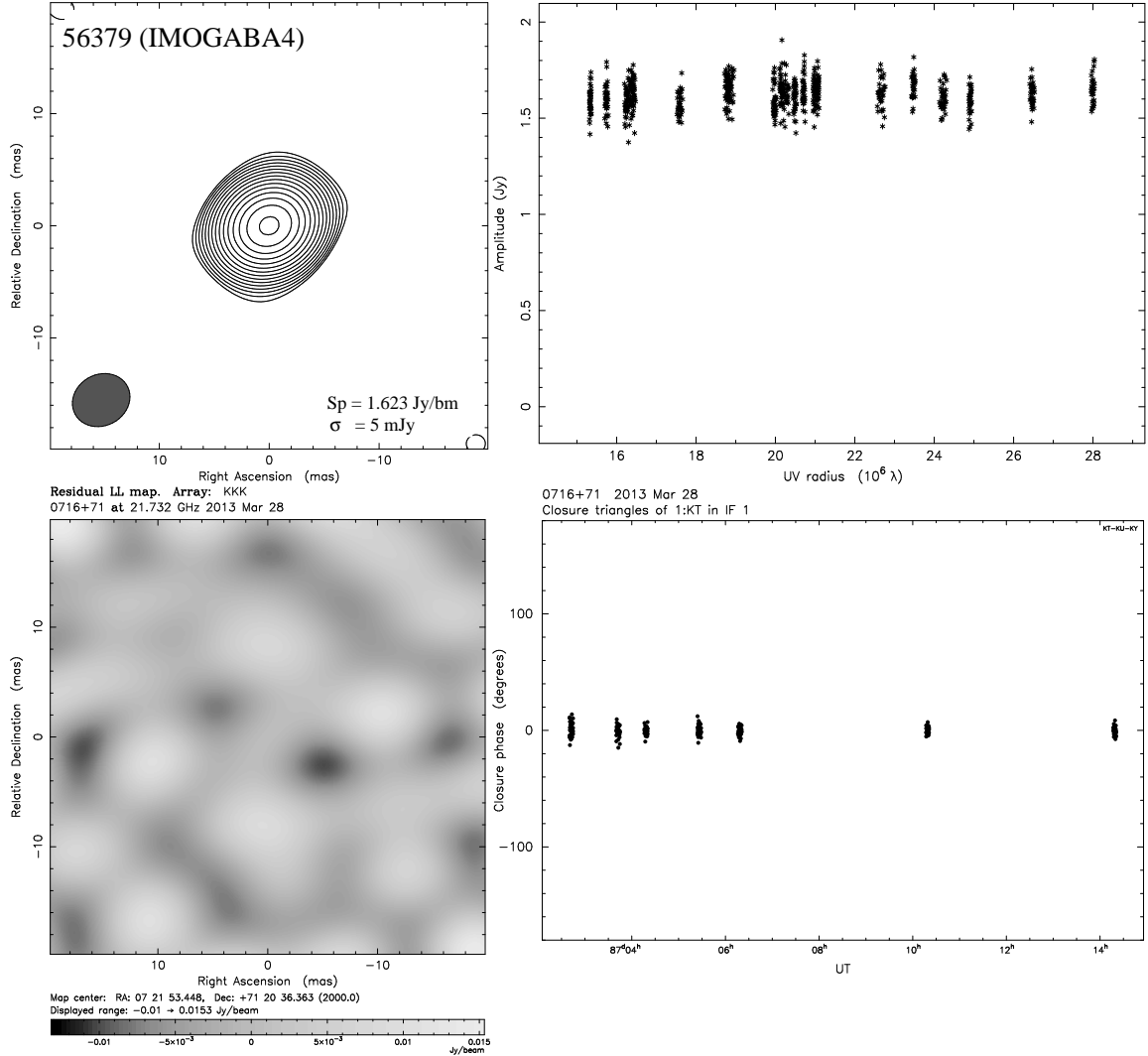}
\caption{
Example of
contour map (upper left),
residual map (lower left),
the visibility amplitude against {\it uv}-radius (upper right),
and the closure phase as function of time (lower right),
at 22 GHz band obtained in March 28, 2013 (MJD 56379).
The axes of the contour and residual maps
show the relative offset from the center of image in milliarcsecond.
Minimum contour level is shown in the lower-right corner of the contour map.
The contours have a logarithmic spacing and are drawn at 
-1, 1, 1.4,...,$1.4^n$ of the minimum contour level. 
In the upper right panel,
the X-axis shows the {\it uv}-distance in $10^6 \lambda$, 
the Y-axis represents the visibility amplitude (correlated flux density)
in Jy.
In the lower right panel,
the X-axis shows the time in UT hour,
and the Y-axis represents the closure phase in degree.
Image parameters of each image are summarized in Table~\ref{t2}.
\label{fig-image}}
\end{figure*}

Figure~\ref{fig-image} shows the representative CLEAN contour maps
of S5 0716+714 for the 22, 43, 86, and 129~GHz bands.
For each VLBI image,
a set of four plots is presented: 
a contour map (top left panel),
a residual map (bottom left panel),
a plot of the visibility amplitude
against the {\it uv}-radius (top right panel),
and a plot of closure phase (bottom right panel).
The contour map is drawn with a relative coordinate
in units of milliarcseconds (mas).
The contour map is centered
on the brightest emission region (VLBI core).
General information about each image is provided in the map,
including 
the FWHM of the restoring beam (the shaded ellipse) in the lower left corner,
and the peak flux density together with the RMS noise level
in the lower right corner of the map.
The contours are drawn with logarithmic spacing,
starting with three times 
the RMS noise level, and increasing as -1, 1, 1.4,...,$1.4^n$
of the lowest contour level.
The plot of the visibility amplitude (top right of each set)
is in units of Jy
for the visibility amplitude (i.e., correlated flux density)
and of $10^6\lambda$, where $\lambda$ is the observing wavelength, 
for the {\it uv}-radius
(i.e., the length of the baseline used to obtain the corresponding visibility
point).

Table~\ref{t2} lists the fitted parameters of the contour maps
presented in Figure~\ref{fig-image}, including the observing frequency band, 
the parameters of the restoring beam (the size of the major axis $B_{\rm maj}$,
the minor axis $B_{\rm min}$ and the position angle of the beam $B_{\rm PA}$),
the total flux $S_{\rm KVN}$,
the peak flux density $S_{\rm p}$,
the off-source RMS $\sigma$,
the dynamic range of image $D$,
and the quality $\xi_{\rm r}$ of the residual 
noise in the image.

Table~\ref{t3} lists the parameters of each model-fit component 
and the measured brightness temperature $T_{\rm b}$. 
The uncertainties of each modelfit parameter are estimated as described
in Section~\ref{imaging-modelfit}.
The upper limits of size $d$, 
and the lower limits of brightness temperature $T_{\rm b}$
are in italic font with flag symbols.

\subsection{Multi-wavelength data\label{data}}

We collected the multi-wavelength flux density data
at 15\,GHz, 43\,GHz, and 230\,GHz of S5 0716+714
obtained as the Owens Valley Radio Observatory (OVRO),
the Very Long Baseline Array (VLBA),
and the Submillimeter Array (SMA), respectively,
from January 2013 to March 2016, as follows.

\subsubsection{OVRO 15\,GHz data}

The 1500 blazar monitoring program at 15\,GHz with
the 40-m radio telescope at the OVRO began in late 2007,
approximately one year before the start of the
Large Area Telescope (LAT) science operation \citep{Richards+11}.
This monitoring is large-scale and fast-cadence monitoring conducted approximately twice a week with
a minimum flux density of 4 Jy and with 3\% typical uncertainty.
S5 0716+714 is one of the published observing sources starting in early 2008.

\subsubsection{VLBA 43\,GHz data}

The VLBA Boston University Blazar monitoring program observes
a total of 36 AGNs including 33 blazars and three radio galaxies
at 43\,GHz mostly once per month \citep{Marscher+08}.
This program is conducted to study the kinematics of the innermost parsec-scale jet
behavior and to determine the location of gamma-ray emission
in the gamma-ray bright AGNs. 
The peak intensity, the polarized intensity, the CLEAN model files,
and the calibrated visibility data files for all observing sources
are available publicly on the VLBA-BU-BLAZAR program website.\footnote{http://www.bu.edu/blazars/VLBAproject.html} 
We obtained the total flux densities of S5 0716+714 from the CLEAN model and the calibrated visibility data
for comparison with those of the iMOGABA project at 43\,GHz.

\subsubsection{SMA 230\,GHz data}

The 230~GHz flux density data were obtained
at the Submillimeter Array (SMA), an eight-element interferometer
located near the summit of Mauna Kea (Hawaii).
S5 0716+714 is included in an ongoing monitoring program
at the SMA to determine the flux densities
of compact extragalactic radio sources that can be used
as calibrators at mm and sub-mm wavelengths~\citep{Gurwell+07}.
The measured source signal strengths were calibrated against
known standards, typically solar system objects
(Titan, Uranus, Neptune, or Callisto).
Data from this program are updated regularly and are available on the SMA
website\footnote{http://sma1.sma.hawaii.edu/callist/callist.html}.

\section{Results and analysis \label{results}}

\subsection{Multi-wavelength light curves}
Figure \ref{fig-multi-lc} shows multi-wavelength light curves of S5 0716+714 obtained during 2013.04 - 2016.16 using the OVRO 40-m radio telescope at 15\,GHz, the KVN at 22-129\,GHz, the VLBA at 43\,GHz, and the SMA at 230\,GHz.

\subsubsection{The 15\,GHz light curve}\label{15lc}
The 15\,GHz light curve is well sampled with a mean cadence of 7 days, which appears to start on a peak or in a middle of a decaying phase of a major flare on 2012.97 with a flux density of $\sim$3.6 Jy and then reaches its minimum level 126 days later on 2013.32.
The light curve shows a prominent second flare with a peak flux density of $\sim$3.6 Jy again on 2013.54, before rapidly decaying to a flux density of $\sim$2 Jy between 2013.54-2013.73. 
This was followed by a more moderate decrease to a flux density of $\sim$1.3 Jy on 2014.22. 
After these two major flares and the mild decline, the 15\,GHz light curve shows relatively small but more rapid flux enhancements than the major flares between 2014.39 and 2015.34, reaching a flux density of $\sim$1.3 Jy.
Another flare, broader in time and not as prominent in flux density, is also seen after 2015.62.
In summary, the 15\,GHz light curve of S5 0716+714 shows two major flares and several small flux enhancements, ranging from 1.0 to 3.6 Jy with a mean of 2.0 Jy.

\subsubsection{The 230 GHz light curve}\label{230lc}
The 230\,GHz light curve is also sampled with an average cadence of 11 days, comparable to the 15\,GHz light curve.
The light curve appears to start (on a flux density level of $\sim$4.5 Jy) immediately after a peak or in a middle of a decaying phase that we can correlate with that of the first flare shown in the 15\,GHz light curve.
The flux density reaches a minimum of $\sim$1.5~Jy, on 2013.24, which is 28 days before of the first minimum (on 2013.32) of the 15\,GHz light curve.
This shows that the brightness of the source reaches its local minimum earlier by $\sim$28 days at 230\,GHz than at 15\,GHz.
The 230\,GHz light curve peaks on 2013.49 with a flux density of $\sim$6 Jy.
If the flare corresponds to the second flare (on 2013.54) of the 15\,GHz light curve, the 230\,GHz light curve appears to lead the 15\,GHz by $\sim$20 days for this flare.
After the major flare, the 230\,GHz light curve shows several smaller but more rapid flux variations and a monotonic flux decrease reaching its minimum on 2014.22, which is similar in time to the end of the monotonic decrease of the 15\,GHz light curve.
After the local minimum, the light curve peaks on 2014.61, showing its most prominent flare with a flux density of 7.2 Jy, which, however, may correspond to the local maximum on 2014.66 of one of the small flux enhancements in the 15\,GHz light curve.
This flare leads to many smaller and more rapid flux enhancements until the end of the light curve.
In summary, the well-sampled 230\,GHz light curve included major flares and minor flux variations ranging from 1.0 to 7.2 Jy with a mean of 3.0 Jy.
For the major flares, the 230\,GHz light curve appears to lead the 15\,GHz light curve by $\lesssim$20 days.

\subsubsection{The 22-129\,GHz light curves}
The 22-129\,GHz KVN light curves were simultaneously obtained at all frequencies but, with a mean cadence of $\sim$30 days and large time gaps due to system maintenance, were not as well sampled as the 15\,GHz and 230\,GHz total intensity light curves.

There were 29 flux measurements at 22\,GHz, 25 at 43\,GHz, 23 at 86\,GHz, and only 11 at 129\,GHz. The relative lack of observations at 129\,GHz was due to system limitations and poor weather.
The VLBA 43\,GHz light curve is also shown to make the light curve denser and for a comparison of the flux density between the VLBA and KVN, indicating that both light curves are good agreement except for a mild difference in a few epochs due to possible calibration uncertainties, for instance, either in the VLBA or in the KVN observations.

Despite their sparseness, the KVN light curves clearly show a trend similar to those of the 15\,GHz and 230\,GHz light curves, in particular for the decaying phase of the first major flare, the monotonic decreasing phase, and the small but rapid flux variations.
The ranges of the flux density are 1.2-3.8 Jy at 22\,GHz, 1.1-3.7 Jy at 43\,GHz, 0.8-3.0 Jy at 86\,GHz, and 0.5-1.9 Jy at 129\,GHz.
The values of the mean flux density are 1.9 Jy, 2.0 Jy, 1.8 Jy, and 1.2 Jy at 22, 43, 86, and 129\,GHz, respectively, indicating that the flux density decreases at the highest frequencies.

\subsection{Flux density variability at 15 and 230\,GHz}\label{sf}
The most well sampled 3-year light curves, i.e., the 15\,GHz and 230\,GHz ones, suggest that S5 0716+714 had major flares with long time scales and minor flux enhancements with shorter time scales during the observing period.
To quantitatively determine the typical time scales of the variability in the observed light curves at 15 and 230\,GHz, we calculated the first-order structure function (SF) analysis as defined in \cite{Heeschen+87}:

\begin{equation}
SF(\tau)=\frac{1}{N}\sum_{i=1}^{N}[f(t_{i})-f(t_{i}+\tau)]^2,
\end{equation} 
where $f(t_{i})$, $\tau$, and $N$ are the flux density at time $t_{i}$, the time lag, and the number of data points, respectively. 

Guided by the apparent different flaring behaviors during the observations, we first divided the light curves into three parts:
part I for 2012.97-2014.22, including two major flares and a monotonic decaying,
part II for 2014.22-2015.34, including several small but rapid flux enhancements,
and part III for 2015.43-2016.12, covering the remaining light curves.
The observed light curves were linearly interpolated for searching for clear peaks in the SF curves.
The SF curves for each part at 15 and 230\,GHz are shown in Figure \ref{fig-sf}.
For both light curves, the SF curve shows a steep rising trend from the shortest time scale and a peak or plateau at the end of the rising trend, corresponding to a typical time scale of the variability in the light curves.

For part I covering the major flares, the SF curves show a steep rising trend with the first peak at the typical time scales of 100 and 103 days (errors are the mean cadence of data points) for 15 and 230\,GHz light curves, respectively, suggesting that the major flares observed at different frequencies can be attributed to a common emission mechanism.
The typical time scales are similar to those reported by \cite{Rani+13}.
The corresponding SF values at the peak are 1.6 and 3.8 for the 15 and 230\,GHz, respectively.
Because the SF value is proportional to the amplitude of the variability~\citep[e.g.,][]{Rani+13}, for the major flare in part I, the amplitude of the variability in the flares is greater at the 230\,GHz than at 15\,GHz.
For parts II and III covering the minor flux enhancements (15\,GHz) and the major flare (230\,GHz), the SF curves show a steep rising trend similar to that of part I, but the peaks in the SF curves of part III at 15\,GHz and part II at 230\,GHz are not clear.
The first peaks in the SF curves of part II at 15\,GHz and part III at 230\,GHz, however, are clear with typical time scales of 43 and 130 days, respectively.
The corresponding SF values at the peak are 0.4 and 5.0 for 15 and 230\,GHz, respectively, implying that there are small but rapid flux enhancements as well as large and slow flares for parts II and III.
Some physical parameters from the variability time scale are described in Section~\ref{para_tvar}.

In order to quantify the amplitude of the variability of the 15\,GHz and 230\,GHz light curves, we estimated the modulation index $m$ following \cite{Kraus+03}
as $m[\%]=100\sigma_{s}/\bar{S}$
where $\sigma_{s}$ and $\bar{S}$ represent
the standard deviation and mean of the flux density $S$, respectively.
The values of $m$ are 24\,\% at 15\,GHz and 45\,\% at 230\,GHz, indicating that S5 0716+714 shows relatively strong variation at 230\,GHz as compared to that at 15\,GHz.

\subsection{Physical parameters from the variability time scale}\label{para_tvar}

\subsubsection{Apparent brightness temperature from variability timescale}\label{tb_var}
Assuming that the flux density variability is intrinsic to the source, the time scale of the variability constrains the size of the emitting region using causality argument.
If the variable component is spherical with Gaussian brightness distribution, we can estimate the brightness temperature using the time scale of the variability as \citep{Rani+13, Fuhrmann+08}
\begin{equation}
T^{app}_{B}=3.47\times{10}^5\cdot\Delta S\left(\frac{\lambda D_{L}}{\tau_{var}(1+z)^2}\right)^2,
\end{equation}
where $\Delta S$ is change of the flux density over time, $\tau_{var}$ is the variability time scale in year, which was obtained from the structure function in Section~\ref{sf}, $\lambda$ is the observing wavelength in cm, $D_{L}$ is the luminosity distance in Mpc, and z is the redshift.
Here we used $D_{L}$=1510 Mpc \citep{Fuhrmann+08}, z=0.127 \citep{Stadnik+14}.
The values of $\Delta S$ are listed in Table~\ref{table:tb}. 
The computed apparent brightness temperatures are $10^{13.8-14.5}$\,K at 15\,GHz, increasing by a factor of $\sim$4 from Part I to Part II,
and $10^{11.5-11.7}$\,K at 230\,GHz, decreasing by a factor of $\sim$2 from Part I to Part III, as listed in Table~\ref{table:tb}. These values at 15\,GHz are
similar to those obtained by~\cite{Rani+13}.

\subsubsection{Doppler factor}\label{doppler_tvar}

We estimated the high apparent brightness temperatures from the variability time scale exceeding inverse Compton (IC) limit of $T_{B}\sim10^{12}$\,K \citep{Kellermann+69} at 15\,GHz.
Assuming that the excessive apparent brightness temperature is caused by relativistic boosting effect of the radiation, we can estimate Doppler factor:
\begin{equation} 
\delta_{var}=(1+z)\left(\frac{T^{app}_{B}}{10^{12}}\right)^{1/(3+\alpha)}
\end{equation}
where $\alpha$ is the spectral index of optically thin emission region (e.g., 86-129\,GHz) and $\alpha$=-0.7 was adopted from the mean spectral index between 86 GHz and 129 GHz.
The calculated Doppler factors are 7-13 at 15\,GHz, increasing from Part I to Part II, and 0.6-0.8 at 230\,GHz, being similar for Part I and III, as listed in Table \ref{table:tb}.
One possibility is that the emitting regions probed at 15 and 230\,GHz may be different.
If this is the case, it can be reasonable that the Doppler factor may not be the same in these regions.
This may also explain the different $T_{B}$ and similar $\tau_{var}$, whilst smaller $\tau_{var}$ would normally be expected for higher frequencies \citep{Rani+13}.

\subsubsection{Size of emitting region}\label{size_tvar}

Assuming again that the variability of the flux density is intrinsic to the source, the size of the emitting region $\theta$ [mas] can be calculated using the Doppler factor and variability time scale \citep[see e.g.,][for details]{Fuhrmann+08}:
\begin{equation}
\theta=0.173\frac{\tau_{var}}{d_{L}}\delta_{var}(1+z),
\end{equation}
where $\delta$ is Doppler factor and $\tau_{var}$ is the time scale of the variability in days.
We obtained the size of emitting region of 0.07-0.09 at 15\,GHz and 0.01 at 230\,GHz which is listed in Table~\ref{table:tb}.
The obtained size of emitting region at 15\,GHz decreases by $\sim$20\,\% from Part I to Part II, and the size at 230\,GHz is similar in Part I and Part III.
The core sizes estimated by the variability time scales are much smaller (up to a factor of 10, for the case of 230~GHz) than the ones extrapolated from the measured model fitting sizes from the KVN data (see Table~\ref{t3}). 
This may be due to the fact that the KVN observations may be affected by
an instrumental beam blending effect
(making the observed core size larger than the true size)
as discussed in Section~\ref{coresize}.

\subsection{Cross-correlation analysis at 15 and 230\,GHz}\label{dcf}

In order to investigate a possible correlation and its corresponding time delay
between the different wavelengths, we selected the 15 and 230\,GHz light curves which are better sampled than the others.
We calculated a cross-correlation
analysis based on the discrete cross-correlation function (DCF) defined
by \cite{Edelson+88} as follows:
\begin{equation}
UDCF_{ij}=\frac{(a_{i}-\bar{a})(b_{j}-\bar{b})}{\sqrt{\sigma^2_{a}-\sigma^2_{b}}},
\end{equation}
where $a_{i}$ and $b_{j}$ are the measurements of data sets $a$ and $b$
for each light curve, $\bar{a}$ and $\bar{b}$ are the means of the data sets, and $\sigma_{a}$ and $\sigma_{b}$ are their corresponding standard deviations of the time series.
The UDCF is binned with an interval $\Delta\tau$
for estimating the DCF
\begin{equation}
DCF(\tau)=\frac{1}{M}\sum{UDCF_{ij}(\tau)}
\end{equation}
for each time lag, where $M$ is the number of data points in the bin. 
We estimated the standard error for each bin with the following equation:
\begin{equation}
\sigma_{DCF}(\tau)=\frac{1}{M-1}\sqrt{\sum[UDCF_{ij}(\tau)-DCF(\tau)]^2}.
\end{equation}

To calculate the DCF, we used binning intervals of 2.2 days, 4.95 days, 4.3 days, and 3.25 days in part I, II, III, and entire period, respectively, which are half the mean cadence of each part, as shown in Figure~\ref{fig-dcf}.
One expects to see a peak in the DCF curve for correlated variations, and the DCF peak time corresponds to the time delay between the two light curves. 
The uncertainty of the time lag was determined using
a model-independent Monte Carlo method~\citep[e.g., see][]{Peterson+98}.
We generated 1000 subsets which were randomly sampled from the light curves,
and calculated a DCF analysis of the subsets, 
determining the time lags from N=1000 simulations.
Based on 1000 time lags,
we obtained a cross-correlation peak distribution (CCPD),
as shown in the right panels of each plot in Figure~\ref{fig-dcf}.
When the distribution of the CCPD is assumed to be a normal distribution, confidence interval of 95\,\%, $\Delta\tau_{95\%}$, can be computed from $\tau-1.96\frac{\sigma}{\sqrt N} <$ $\tau_{95\,\%}$ $< \tau+1.96\frac{\sigma}{\sqrt N}$. Here, N is number of simulation, $\tau$ is time lag with negative values meaning 230\,GHz leads, and $\sigma$ is standard deviation of the CCPD.
The values of $\tau$ are -17.36 days, -13.80 days, 44.49 days, and -15.88 days in part I, II, III, and entire period, respectively. The values of $\sigma$ are 4 in part I, 5 in part II, 44 in part III, and 8 days in entire period.
The confidence interval for the part I is -17.59 $< \tau_{95\,\%} <$ -17.12, for the part II is -14.14 $< \tau_{95\,\%} <$ -13.46, and for the part III is 42.79 $< \tau_{95\,\%} <$ 46.19 for the entire period is -16.38 $< \tau_{95\,\%} <$ -15.38.

The time lag between 15 and 230\,GHz for the entire period of the light curves
is $-16\pm8$ days (top left panel in Figure~\ref{fig-dcf}),
implying that the 230~GHz light curve leads the 15~GHz light curve.
For part I of the light curves,
the time lag is also $-17\pm4$ days, similar to the time differences
of 20 days estimated above for the peaks in the light curves at 15 and 230\,GHz (see \ref{230lc}),
confirming that the 230~GHz light curve leads the 15~GHz light curve for major flares.
The DCF time lag for part II is $-14\pm5$ days, close to the time differences
of 17 days for the peaks in the 15 and 230\,GHz light curves (as discussed in Section~\ref{230lc}),
indicating that the 230~GHz light curve also leads the 15~GHz light curve for small flux enhancements.
In this part, the multiple peaks of the DCF seem to be caused by correlation with the consecutive flares.  
Contrary to parts I and II,
for part III, the determined time lag is $44\pm27$ days.
This result appears artificial
because there is no well sampled measurement for the beginning portion of part III in the 230\,GHz light curve and hence no corresponding measurements of the flux enhancements in the 15\,GHz light curve.

In summary,
the flares in the 230\,GHz light curves lead
those of the 15\,GHz light curve with a time lag of $-14\sim-17$ days,
except for part III.
The time lag between flux enhancements at multi-frequency can be regarded by the opacity effect in the core region.
Therefore, the postion of the VLBI core at 15\,GHz with respect to 230\,GHz can be drived by using the relation of $r_{core}=\mu\cdot\tau$~\citep{Kudryavtseva+11}.
$\mu$ and $\tau$ correspond proper motion of the source in mas yr$^{-1}$ and the time lag derived from the DCF in year, respectively.
Here, we adopted $\mu$=2.258 mas~yr$^{-1}$ \citep{Lister+16} and derived $r_{core}=0.09-0.1$ mas.
This value shows a good agreement with a core position of $r_{core}=0.10\pm0.03$ mas between 15 and 43\,GHz measured from the VLBI by~\citet{Marti-Vidal+16}.

\subsection{Brightness temperature}\label{tb}

Brightness temperatures of the VLBI cores obtained from our VLBI measurements vary in time as shown in Figure~\ref{fig-tb} and summarized in Table~\ref{t3}. 
The observed brightness temperatures are in the ranges of
$10^{8.3-10.3}$~K,
$10^{9-10.8}$~K,
$10^{9.2-10.6}$~K,
and $10^{9.1-10.6}$~K
at 22, 43, 86, and 129\,GHz, respectively,
including the lower limits of the brightness temperature,
which were obtained for the unresolved cores (see Section~\ref{imaging-modelfit}).
The mean brightness temperatures are
$10^{8.8}$ K at 22\,GHz,
$10^{9.7}$ K at 43\,GHz,
$10^{10.1}$ K at 86\,GHz, and
$10^{10.2}$ K at 129\,GHz,
excluding the lower limits of the brightness temperatures,
implying that the observed core brightness temperatures of S5 0716+714
become higher at higher frequency.

The observed brightness temperatures are lower than those
obtained with the Very Long Baseline Array at 43 and 86~GHz
by factors of about 10-200~\citep{Hodgson+15},
and with the Global Millimeter VLBI Array at 86~GHz
by factors of about 4-20~\citep{Lee+08}.
These results are also lower than the brightness temperature
derived from the source variability at 86~GHz
by orders of up to a factor of 3~\citep{Rani+13}.
The difference may be due to the spatial resolution difference between
the KVN and the global VLBI arrays,
and to the instrumental beam blending
effect (making the observed core size larger than the true size)
as discussed in Section~\ref{coresize} and \cite{Lee+16}.
The observed brightness temperatures are strongly correlated with
the observed core size with correlation coefficient $r$ up to -0.74, whereas those are not with the CLEAN flux density with $r$ as small as 0.15
as shown in Figure~\ref{fig-tb-corr}, implying that the instrumental
beam blending effect may play an important role in the brightness temperature
estimation.	

\subsection{Spectral index}\label{spectra1}
Simultaneous multi-frequency observations with
the KVN allow us to study the variation of 
the spectral index $\alpha$
($S_{\nu} \propto \nu^{\alpha}$,
where $\nu$ is the observing frequency, and
$S_{\nu}$ is the flux density).
In order to estimate the spectral index not only with KVN measurements
but also with non-KVN measurements,
we selected the non-KVN (i.e., OVRO) flux measurements
when they were obtained
closely to the KVN observations in time
within one day (quasi-simultaneous).
We found that there are six quasi-simultaneous flux measurements
of OVRO, with regard to the KVN observations,
as shown in Figure~\ref{fig-multi-sindex}.
In this figure, 1-${\sigma}$ error bars are plotted. The errors were estimated as
\begin{equation}
\sigma_{i,j}=\frac{\sqrt{(\sigma_{i}^2/S_{i}^2)+(\sigma_{j}^2/S_{j}^2)}}{\rm log(\nu_{i}/\nu_{j})}
\end{equation}
where, $\sigma$, $S$, and $\nu$ are measurement error of the flux density, the flux density, and the observing frequency, respectively. 

The spectral indices vary in time along $\alpha=0$ as summarized in Table~\ref{table:sindex_range}.
The spectral indices for most of the frequency pairs
are $-0.5<\alpha<0.5$. 
For the 43-86~GHz, 43-129~GHz, and 86-129~GHz,
the spectral index became steeper than $-$0.5, i.e., optically thin.
The mean (and standard deviation) values of the spectral index
are 0.13 (0.11) at 15-22\,GHz,
0.16 (0.14) at 15-43\,GHz,
0.06 (0.19) at 15-86\,GHz,
0.06 (0.17) at 22-43\,GHz,
$-$0.09 (0.16) at 22-86\,GHz,
$-$0.27 (0.13) at 22-129\,GHz,
$-$0.25 (0.19) at 43-86\,GHz,
$-$0.44 (0.17) at 43-129\,GHz,
and $-$0.69 (0.32) at 86-129\,GHz,
indicating that the spectral index becomes steeper at pairs at higher frequencies.

By comparison of Figure~\ref{fig-multi-lc} and \ref{fig-multi-sindex}, we see that the spectral indices follows similar trend to those of the observed flux densities.
In order to investigate a correlation between the spectral indices
and the flux densities of the observing frequencies,
we estimated the Pearson correlation coefficient $r$.
Figure~\ref{fig-flux-sindex} shows 
the spectral index as a function of the flux density at lower (left panels)
and higher (right panels) frequency for 11 out of 29 epochs
which yield the flux densities in all four frequency bands.
Generally, the correlations between the spectral index and the flux density
for each pair of frequencies are stronger (i.e., $r>0.5$) at a higher frequency,
implying that the variation of the spectral index is more correlated with the flux density variation at a higher frequency than that at a lower frequency.

\subsection{Source spectra}\label{spectral2}
The simultaneous multi-frequency VLBI images obtained at four frequency bands (22, 43, 86, and 129\,GHz) during such a long period of time are the first results for S5 0716+714, leading to the study of mas-scale spectral information without any uncertainty due to source variability.
We used the CLEAN flux densities obtained on mas scales to investigate the spectral properties of the source.

Two models were used for fitting to the CLEAN flux density data:
the power law and the curved power law, as given by
\begin{equation}
\label{eqn:plaw}
S \propto \nu^{+\alpha},
\end{equation}
where $S$ is the CLEAN flux density in Jansky,
$\nu$ is the observing frequency in GHz,
$a$ is constant in Jansky,
and $\alpha$ is the spectral index for the power law, and
\begin{equation}
\label{eqn:claw}
S=c_1 \left(\frac{\nu}{\nu_{\rm r}}\right)^{\alpha +c_2 {\rm ln}(\nu/\nu_{\rm r})},
\end{equation}
where
$c_1$ (in Jansky) and $c_2$ are constant,
$\nu_{\rm r}$ is a reference frequency,
and $\alpha$ is the spectral index at $\nu_{\rm r}$.

Initially, the power law is fitted to the data.
When the power law does not fit well the data due to the spectral curvature,
a curved power law is used.
For 8 out of 28 epochs,
the power law was fitted to CLEAN flux measurements
which were obtained at only two frequencies.
The curved power law was well fitted to the CLEAN flux measurements
for the remaining 20 epochs when the CLEAN flux measurements
were available at more than three frequencies.
We determined the peak flux density and turnover frequency by fitting the curved power-law with $\alpha=0$.
The CLEAN flux spectra of 28 epochs are shown
in Figure~\ref{fig-spind} with their best-fit model and
fit parameters in table~\ref{table:nu_c,s_m,bfield}. Shown are the spectral index $\alpha$ for the power law, and
the turnover frequency $\nu_{\rm c}$ and peak flux density $S_{\rm m}$ at $\nu_{\rm c}$ obtained from the curved power law. 

\section{Discussion \label{discussion}}

\subsection{Turnover frequency}\label{turnover}
From the spectral model fitting, we were able to obtain $\nu_{\rm c}$ and $S_{\rm m}$ for 20 epochs,
as summarized in Table~\ref{table:nu_c,s_m,bfield} and
as shown in panels (1) and (2) of Figure~\ref{fig-multi-curve}, where we plot $S_{\rm m}$ and $\nu_{c}$ as a function of the observing time in order to investigate the variation of $\nu_{\rm c}$ and $S_{\rm m}$ in the time domain.
We performed a Monte Carlo simulation with generating N=1000 spectra to obtain statistically meaningful values for the turnover frequency and the peak flux density as well as their uncertainties~\citep[see e.g.][]{Fromm+15}.
$\nu_{\rm c}$ is in the range of 21\,GHz-69\,GHz
with a mean turnover frequency $\bar{\nu}_{\rm c}$ of 37.7\,GHz.
The standard deviation of the turnover frequency $\sigma_{\nu_{\rm c}}$ is 14~GHz.
$S_{\rm m}$ ranges from 1.21 Jy to 3.89 Jy
with a mean peak flux density $\bar{S}_{m}$ of 2.3~Jy
and corresponding standard deviation $\sigma_{S_{\rm m}}$ of 0.74~Jy.
It is apparent that $S_{\rm m}$ and $S_{\nu}$ are correlated each other
as well as with the KVN light curves.

In order to study the evolution of the spectra over the time,
we selected three periods of time in the KVN light curves:
period (a) for 2013.04-2013.24,
period (b) for 2013.79-2014.16,
and period (c) is 2015.81-2015.99,
which are relatively well sampled periods
and which show a decaying trend.
Figure~\ref{fig-nuc-sm} shows the evolution of
the peak flux density $S_{\rm m}$
and the turnover frequency $\nu_{\rm m}$ for periods (a), (b), and (c).
In this case, $S_{\rm m}$ increases
as $S_{\rm m}\propto \nu_{\rm  m}^{\epsilon_{\rm adiabatic}}$
with $\epsilon_{\rm adiabatic}=1.8$ for period (a), 
$\epsilon_{\rm adiabatic}=0.6$ for period (b), 
and $\epsilon_{\rm adiabatic}=0.9$ for period (c), 
which is consistent to the adiabatic stage
(e.g., $\epsilon_{\rm adiabatic}=0.69$ in \cite{Marscher+85}
$\epsilon_{\rm adiabatic}=0.77$ in \cite{Fromm+11},
and $\epsilon_{\rm adiabatic}=2$ for R8 flare of S5 0716+714
in \cite{Rani+13})
in the spectral evolution model of \cite{Marscher+85},
who discuss the evolution of a propagating shock wave
in a steady state jet (or the shock-in-jet model).
This may indicate that
the flares are produced by a shock in the relativistic jet
of S5 0716+714.

The correlations of the flux density at observing frequencies $S_{\nu}$
with $S_{\rm m}$ and $\nu_{\rm c}$ were investigated
using the estimated linear correlation coefficient $r$, as
shown in Figure~\ref{fig-flux-sm_nuc}.
The linear correlation coefficients $r$ between
$S_{\nu}$ and $S_{\rm m}$ are 0.94 at 22\,GHz, 1.0 at 43\,GHz, 0.91 at 86\,GHz and 0.89 at 129\,GHz,
indicating that the peak flux density is strongly
correlated with the flux density and shows the best correlation with 
the 43~GHz flux density.
The linear correlation coefficients $r$ between
$S_{\nu}$ and $\nu_{\rm c}$ are 0.01 at 22\,GHz, 0.32 at 43\,GHz, 0.62 at 86\,GHz and 0.57 at 129\,GHz,
suggesting that the turnover frequency is less correlated with
the flux density than with the peak flux density.
The turnover frequency is most strongly correlated with the flux density
at 86\,GHz, indicating that the variation of the emission in optically thin region
is very closely related to the change in the physical properties of synchrotron self-absorption.

\subsection{Deconvolved core size}\label{coresize}
We obtained the core sizes deconvolved
from the synthesized beams of KVN using a two-dimensional Gaussian model-fit,
as summarized in Table~\ref{t2}.
The ranges of the deconvolved core sizes are 0.1-1.5 mas at 22\,GHz,
0.09-0.6 mas at 43\,GHz,
0.09-0.4 mas at 86\,GHz,
and 0.09-0.3 mas at 129\,GHz,
including the upper limits of the size
(the downward arrows in panel (5) in Figure~\ref{fig-multi-curve}).
The mean (standard deviation) values of the deconvolved core sizes are 0.9 (0.2) mas, 0.3 (0.1) mas, 0.16 (0.08) mas, and 0.15 (0.07) mas
at 22, 43, 86, and 129\,GHz, respectively,
excluding the upper limits of the size.  

The deconvolved core sizes at 22\,GHz and 43\,GHz
from the iMOGABA observations are $\sim$10 times larger
than those from the VLBA observations at 22\,GHz~\citep[][$\theta\sim0.1$~mas]{Bach+05}
and at 43\,GHz~\citep[][$\theta\sim0.04$~mas]{Rani+15}, respectively, for S5 0716+714.
This is mainly attributed to the relatively large synthesized beams
of the KVN (6 mas at 22~GHz and 3 mas at 43~GHz) as compared to those of the VLBA (0.3 mas at 22~GHz and 0.15 mas at 43~GHz),
which may lead to an instrumental beam blending effects: the deconvolved core region is
diluted by jet emissions coming from regions downstream of the $\tau=1$ surface~\citep{Lee+16}.

As shown in panel (5) of Figure~\ref{fig-multi-curve},
the deconvolved core size was found to vary in time,
which may be due to several factors, such as
(a) errors when deconvolving the core with two-dimensional Gaussian models,
(b) the variation of the viewing angle of the core emission region,
(c) the ejection of a new jet component still located in the blended core.

The fitting errors are
0.010 mas at 22\,GHz,
0.016 mas at 43\,GHz,
0.020 mas at 86\,GHz,
and 0.024 mas at 129\,GHz,
which are much smaller than the standard deviations (0.07-0.2 mas)
of the deconvolved core sizes at 22, 43, and 86\,GHz,
ruling out fitting errors due to time variations of the core size.
The variation of the viewing angle from $\psi_{1}$ to $\psi_{2}$ may cause a fractional change of the deconvolved core size via $|\Delta d|/d=|cos\psi_{2}-cos\psi_{1}|$ which may be at most 0.015 (1.5\,\%) for a large change of the viewing angle from $\psi_{1}\sim1^{\circ}$ to $\psi_{2}\sim10^{\circ}$,
as reported in \cite{Rani+15}.
This value is much smaller than the fractional change (20\,\%-50\,\%) of the deconvolved core size obtained from the iMOGABA observations.

In the case of the ejection of a new jet component,
the variation of the deconvolved core size
may depend on the jet angular speed $\mu$.
The mean angular speed $\bar{\mu}$ of new jet components
of S5 0716+714 is as large as $\mu\sim$1~mas yr$^{-1}$~\citep[see][]{Rani+15}.
When a new jet component emerges from the core region and is blended with the core, the blended core size $d$ may increase as $\Delta d \propto \mu$.
If the new jet component is resolved from the core,
the deconvolved core size may rapidly decrease.
The standard deviations of the deconvolved core sizes in the iMOGABA observations
are larger than 0.1 mas (at 22, 43, and 86\,GHz), corresponding to the angular movement of
the new jet component of S5 0716+714 in one month
(i.e., the observing cadence of the iMOGABA).
Therefore, the time variation in the deconvolved core size
may be partly attributed to the instrumental beam blending effect
and a newly ejected jet component still located in the blended core region.
We may find, for example,that the deconvolved core size increases
from 0.1 mas to 0.4 mas at 86\,GHz
during 2015.91-2016.11
(i.e., $\Delta d\sim1.6$~mas yr$^{-1}$).

\subsection{Magnetic field}\label{bfield}
The turnover frequency $\nu_{c}$ and peak flux density $S_{\rm m}$
determined by the spectral model fitting (see Section~\ref{turnover})
enable us to constrain the magnetic field
in the emission region dominating the flux density at the turnover frequency,
assuming that the emission region is a compact
and hence homogeneous synchrotron self-absorption (SSA)
region, following \cite{Marscher+83} and \cite{Hodgson+17} as  
\begin{equation}\label{eq:b-field}
B=10^{-5}b(\alpha)S^{-2}_{\rm m}\theta^4\nu_{c}^5\left(\frac{\delta}{1+z}\right)^{-1}~~~(G),
\end{equation} 
where $b(\alpha)$ is a parameter which depends on optically thin spectral index ranging from 1.8 to 3.8~\citep[see Table 1 in][]{Marscher+83},
$S_{\rm m}$ is the peak flux density in Jy,
$\theta$ is the angular size (mas) of the emission region,
$\nu_{\rm c}$ is the turnover frequency in GHz,
$\delta$ is the Doppler factor,
and $z$ denotes the redshift.
We should note that Equation~(\ref{eq:b-field}) is updated from
that in \cite{Marscher+83} with the $\left(\frac{\delta}{1+z}\right)$
factor being raised to the $-1$ power rather than $+1$ power,
since the VLBI core of this source is in a steady state rather than
evolving in time (A. Marscher, private communication).

Given that the deconvolved core size from the iMOGABA observations
may include the instrumental beam blending effect,
we used $\theta = 0.04$~mas obtained from the VLBI observations,
and assumed $\delta = 7$, following \cite{Rani+13}.
Using $b(\alpha)$ = 2.92,
$S_{\rm m}$ = 1.2-3.9 Jy,
and $\nu_{\rm c}$ = 21.1-69.4\,GHz,
we obtained the upper limits of the magnetic field $B$ of 
$\le5$~mG during the observing period,
as summarized in Table~\ref{table:nu_c,s_m,bfield},
yielding a weighted mean of
$1.0\pm0.6$~mG for higher turnover frequencies (e.g., $>$45~GHz).
This result is consistent with those in \cite{Rani+13} when applying equation~(\ref{eq:b-field}).

\section{Summary}\label{summary}

Three years of multi-frequency VLBI monitoring observations of S5 0716+714
using KVN at 22, 43, 86, and 129\,GHz, together with 15 and 230~GHz total intensity data, enabled us to find the following conclusions:

\begin{itemize}

\item 	S5 0716+714 is compact on the mas scale at 22-129~GHz under the KVN resolution, showing only a VLBI core component. During the observing period of time,
there were several major flares with peak flux densities 
of $\sim$4~Jy at 15~GHz and $\sim$7~Jy at 230~GHz.

\item 	The typical time scales of the flares are 43-100 days at 15\,GHz and 100-130 days at 230\,GHz.  
	The 230\,GHz light curve leads the 15\,GHz one with a time-lag of -16$\pm$8 days for these flares.

\item 	The two-frequency spectral indices are in the range of -0.5 to 0.5 and
	are steeper at higher frequency pairs with $\alpha\le-1.5$.
	The variation in the two-frequency spectral index is more correlated
	with the flux density at a higher frequency, i.e., in the case of optically thin emission.

\item	The spectrum of the source obtained from the simultaneous multifrequency VLBI observations
	at 22-129\,GHz has a spectral break (i.e., turnover frequency)
	and varies in time. The turnover frequency varies from 21~GHz to 69~GHz
	with its peak flux density ranging from 1~Jy to 4~Jy.

\item	The evolution of the turnover frequency and the peak flux density can be well described
	by the shock-in-jet model~\citep{Marscher+85}, especially as an adiabatic stage
	during the decreasing phase of the flux density.

\item	The deconvolved core sizes obtained from the two-dimensional Gaussian modelfit
	are $\sim$10 times larger at 22 and 43\,GHz than
	previous VLBI measurements~\citep[e.g.,][]{Bach+05,Rani+15}
	mainly due to instrumental beam blending effects.
	We partially explain the time variation of the deconvolved core size
	(e.g., at 86\,GHz) by the ejection of new jet components.

\item 	By assuming a synchrotron self-absorption model for the relativistic jet emission in S5 0716+714,
	we estimated the magnetic field strength in the mas emission region
	to be $\le5$~mG during the observing period, yielding 
	a weighted mean of $1.0\pm0.6$~mG for higher turnover frequencies
	(e.g., $>$45~GHz).

\end{itemize}

\acknowledgments
We would like to thank the anonymous referee for important comments
and suggestions which have enormously improved the manuscript.
We also thank Thomas Krichbaum and Bindu Rani
for careful reading and kind comments to the manuscript.
We are grateful to all staff members in KVN
who helped to operate the array and to correlate the data.
The KVN is a facility operated by
the Korea Astronomy and Space Science Institute.
The KVN operations are supported
by KREONET (Korea Research Environment Open NETwork)
which is managed and operated
by KISTI (Korea Institute of Science and Technology Information).
This research has made use of data from the OVRO 40-m monitoring program (Richards, J. L. et al. 2011, ApJS, 194, 29) which is supported in part by NASA grants NNX08AW31G, NNX11A043G, and NNX14AQ89G and NSF grants AST-0808050 and AST-1109911.
This work was supported by the National Research Foundation of Korea(NRF) grant funded by the Korea government(MSIP) (No. NRF-2016R1C1B2006697).
S.S.K. was supported by the National Research Foundation grant funded by the Ministry of Science, ICT and Future Planning of Korea (NRF-2014R1A2A1A11052367).
The Submillimeter Array is a joint project between the Smithsonian Astrophysical Observatory and the Academia Sinica Institute of Astronomy and Astrophysics and is funded by the Smithsonian Institution and the Academia Sinica.
We are grateful to Alan Marscher for very fruitful comments on the magnetic field estimation.


\bibliographystyle{aasjournal}
\bibliography{imogaba0716_arxiv}

\setcounter{figure}{1}
\begin{figure*}[!t]
\epsscale{2}
\plotone{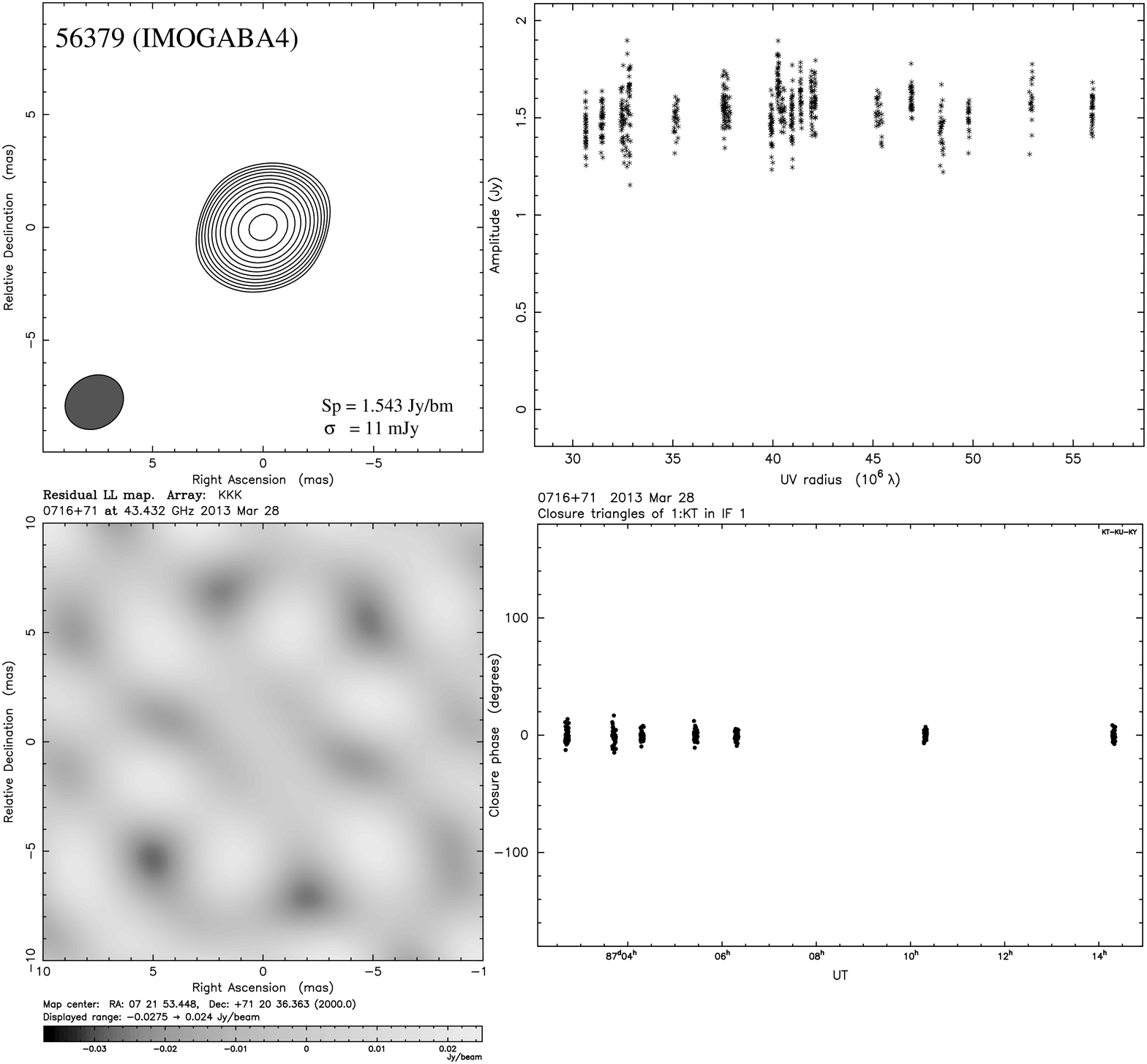}
\caption{{\it Example of
contour map (upper left),
residual map (lower left),
the visibility amplitude against {\it uv}-radius (upper right),
and the closure phase as function of time (lower right),
at 43 GHz band obtained in March 28, 2013 (MJD 56379) (continued)}}
\end{figure*}
\setcounter{figure}{1}
\begin{figure*}[!t]
\epsscale{2}
\plotone{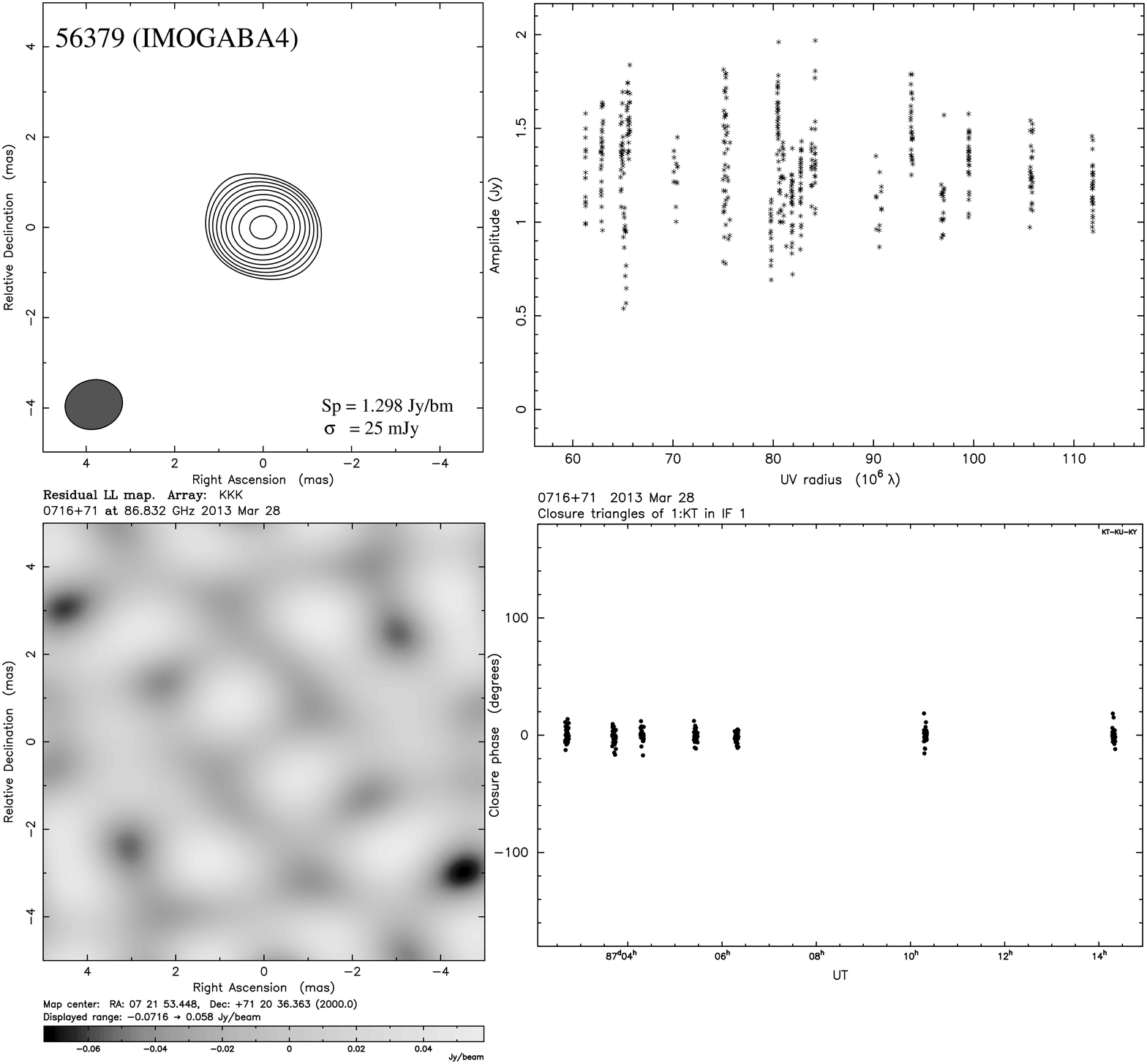}
\caption{{\it Example of
contour map (upper left),
residual map (lower left),
the visibility amplitude against {\it uv}-radius (upper right),
and the closure phase as function of time (lower right),
at 86 GHz band obtained in March 28, 2013 (MJD 56379) (continued)}}
\end{figure*}
\setcounter{figure}{1}
\begin{figure*}[!t]
\epsscale{2}
\plotone{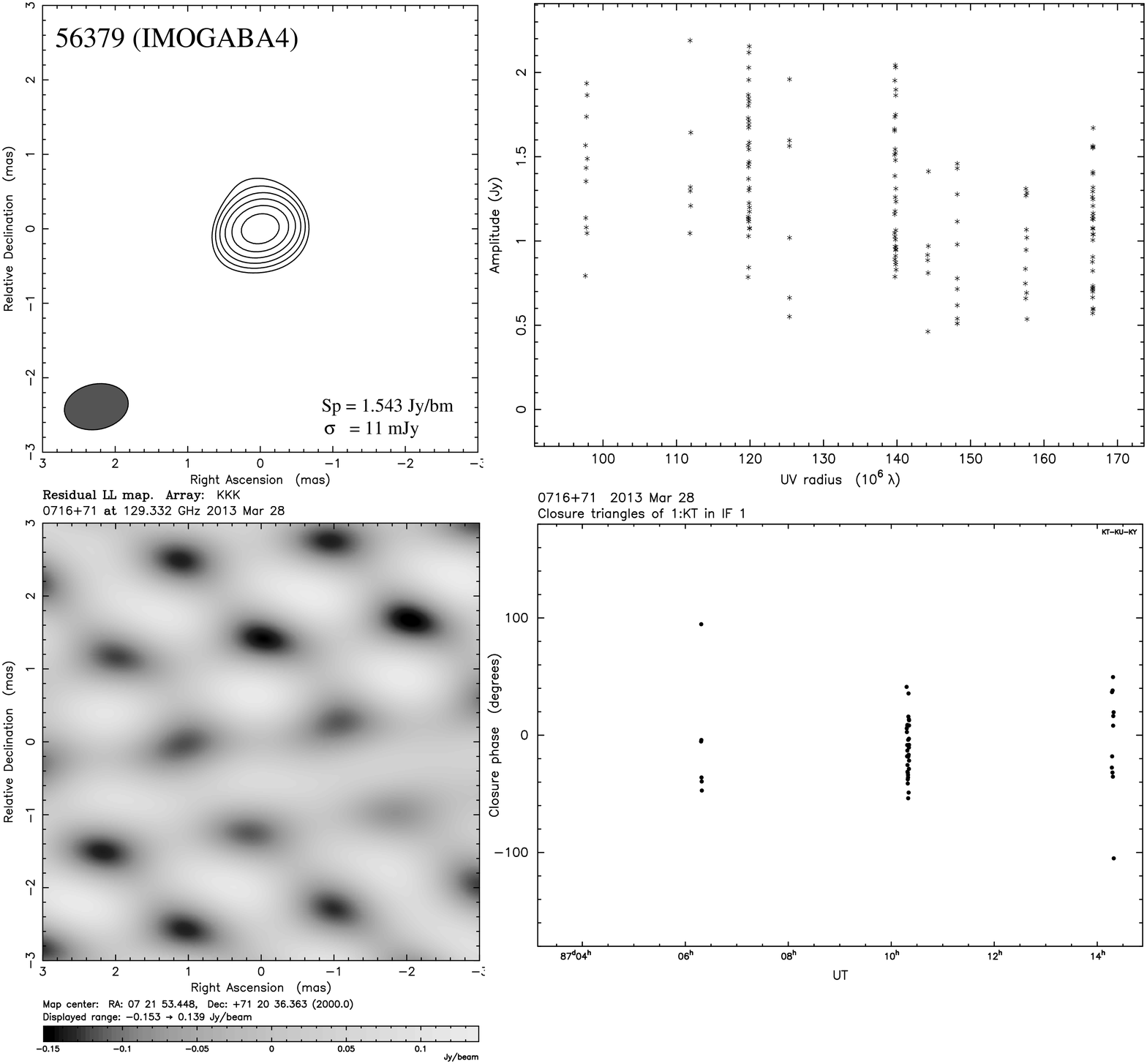}
\caption{{\it Example of
contour map (upper left),
residual map (lower left),
the visibility amplitude against {\it uv}-radius (upper right),
and the closure phase as function of time (lower right),
at 129 GHz band obtained in March 28, 2013 (MJD 56379)}}
\end{figure*}

\clearpage
\begin{figure*}[!t]
\epsscale{1.9}
\plotone{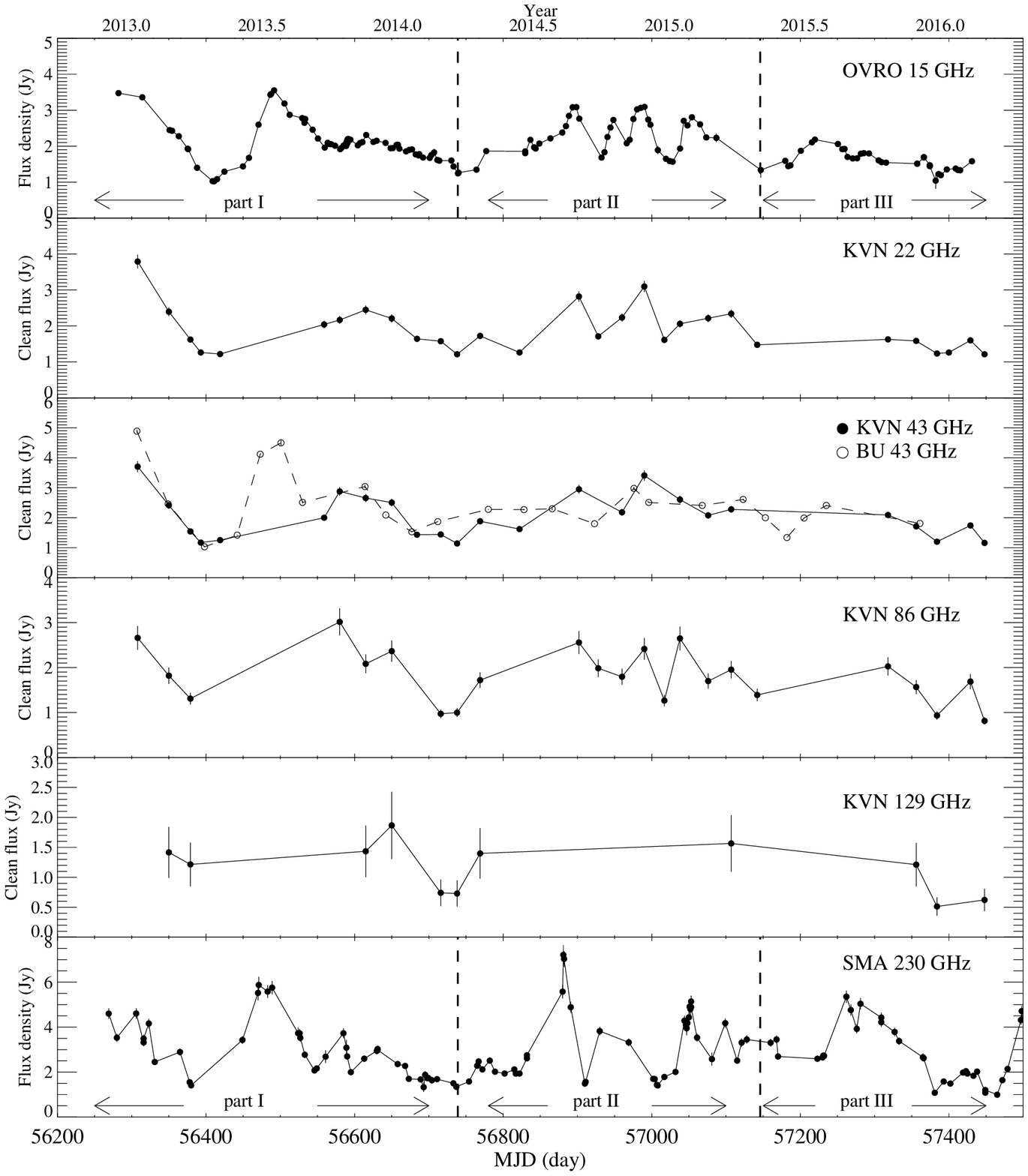}
\caption{Light curves of S5 0716+714 from radio to mm wavelength observed during 2013.04-2016.16. 
\label{fig-multi-lc}}
\end{figure*}

\begin{figure*}[!t]
\epsscale{1.9}
\plotone{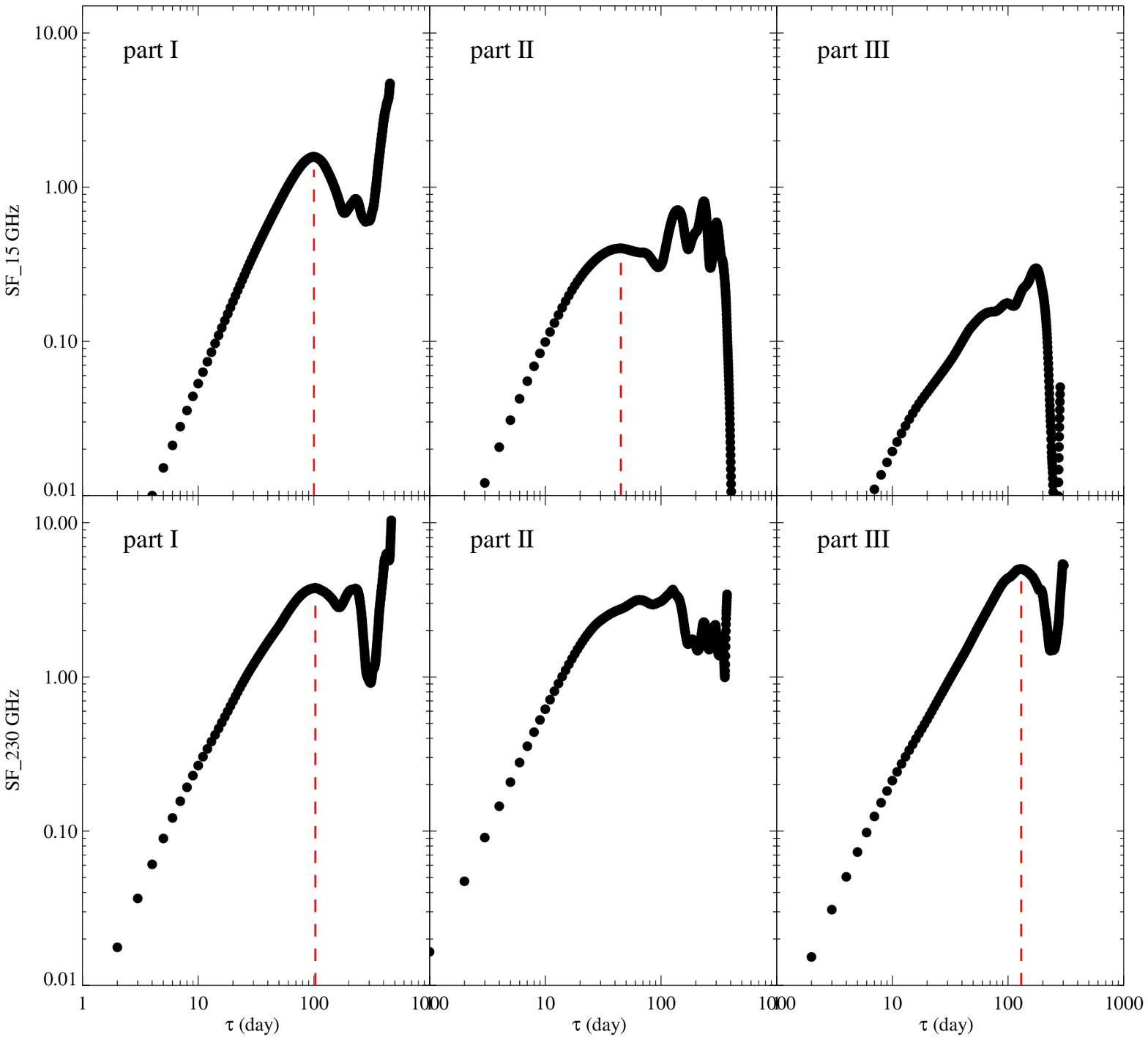}
\caption{Structure function of S5 0716+714 for the parts I, II, and III of the 15\,GHz (top) and 230\,GHz (bottom) light curves. 
\label{fig-sf}}
\end{figure*}
\clearpage
\begin{figure*}[!t]
\epsscale{2}
\plotone{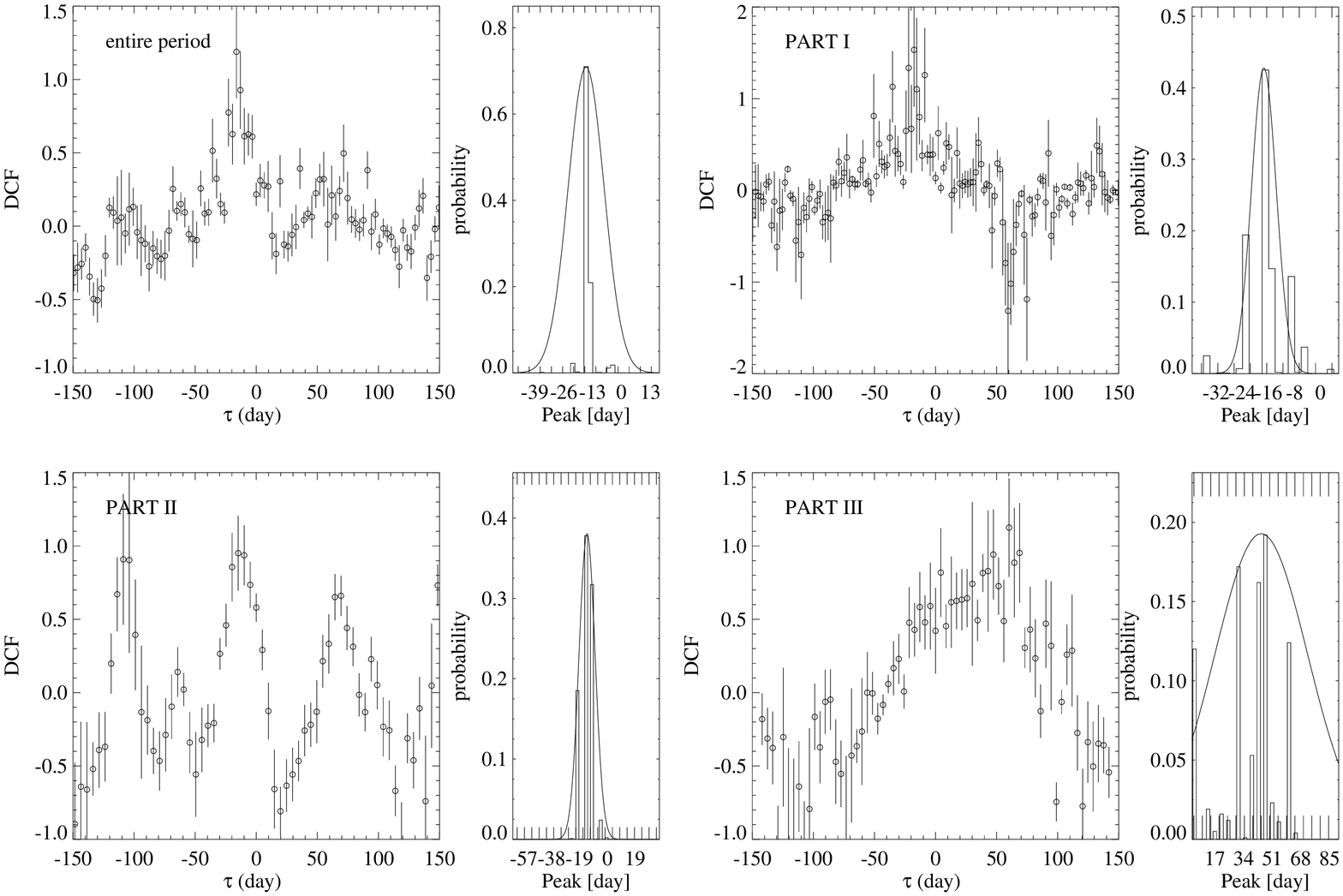}
\caption{Cross-correlation functions and the corresponding cross-correlation peak distributions (CCPD)
of the light curves at 15 and 230\,GHz in S5 0716+714
for the entire period (top left), part I (top right), part II (bottom left),
and part III (bottom right),
and the corresponding cross-correlation peak distributions (CCPD).
\label{fig-dcf}}
\end{figure*}

\begin{figure*}[!t]
\epsscale{1.9}
\plotone{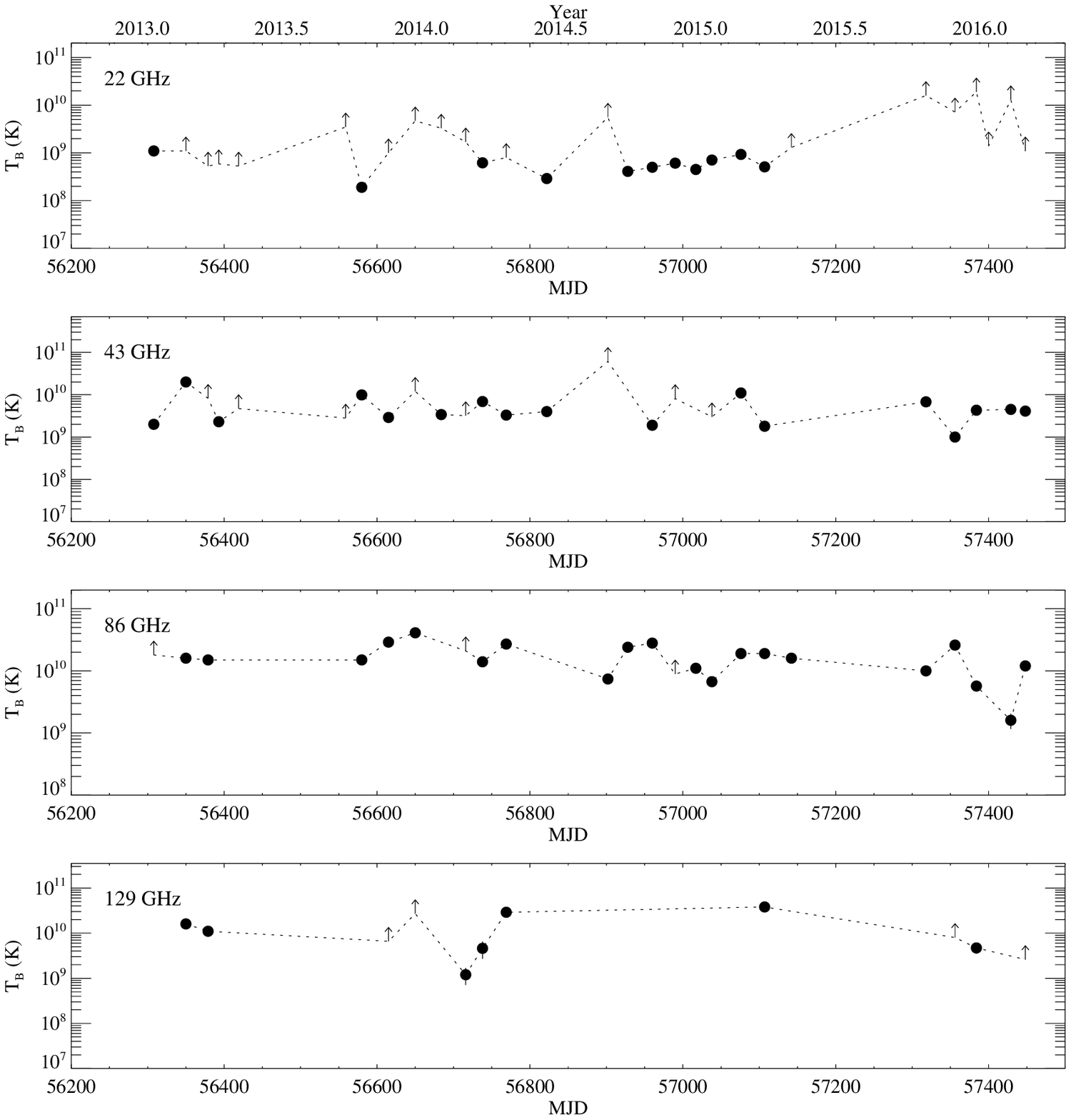}
\caption{ Brightness temperatures of the VLBI core of S5 0716+714
observed in the 22~GHz, 43~GHz, 86~GHz, and 129~GHz bands, from top to bottom. Arrows indicate the lower limits.
\label{fig-tb}}
\end{figure*}

\begin{figure*}[!t]
\epsscale{1.9}
\plotone{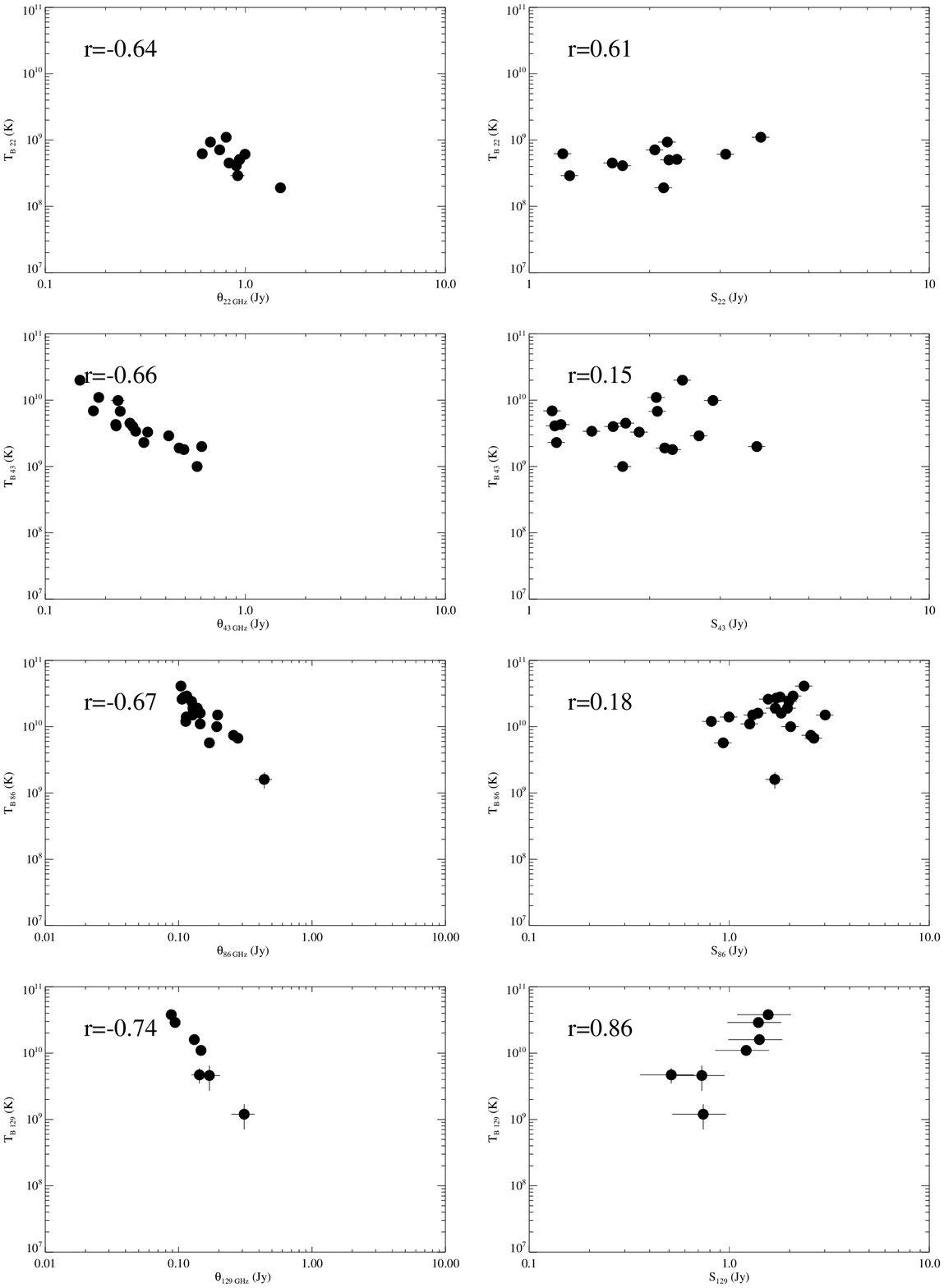}
\caption{Brightness temperature as a function of deconvolved core size (left panels) and CLEAN flux density (right panels) for the different observing frequencies. The Pearson coefficient for a potential correlation is shown at each panel at the upper, left corner.
\label{fig-tb-corr}}
\end{figure*}

\begin{figure*}[!t]
\epsscale{1.8}
\plotone{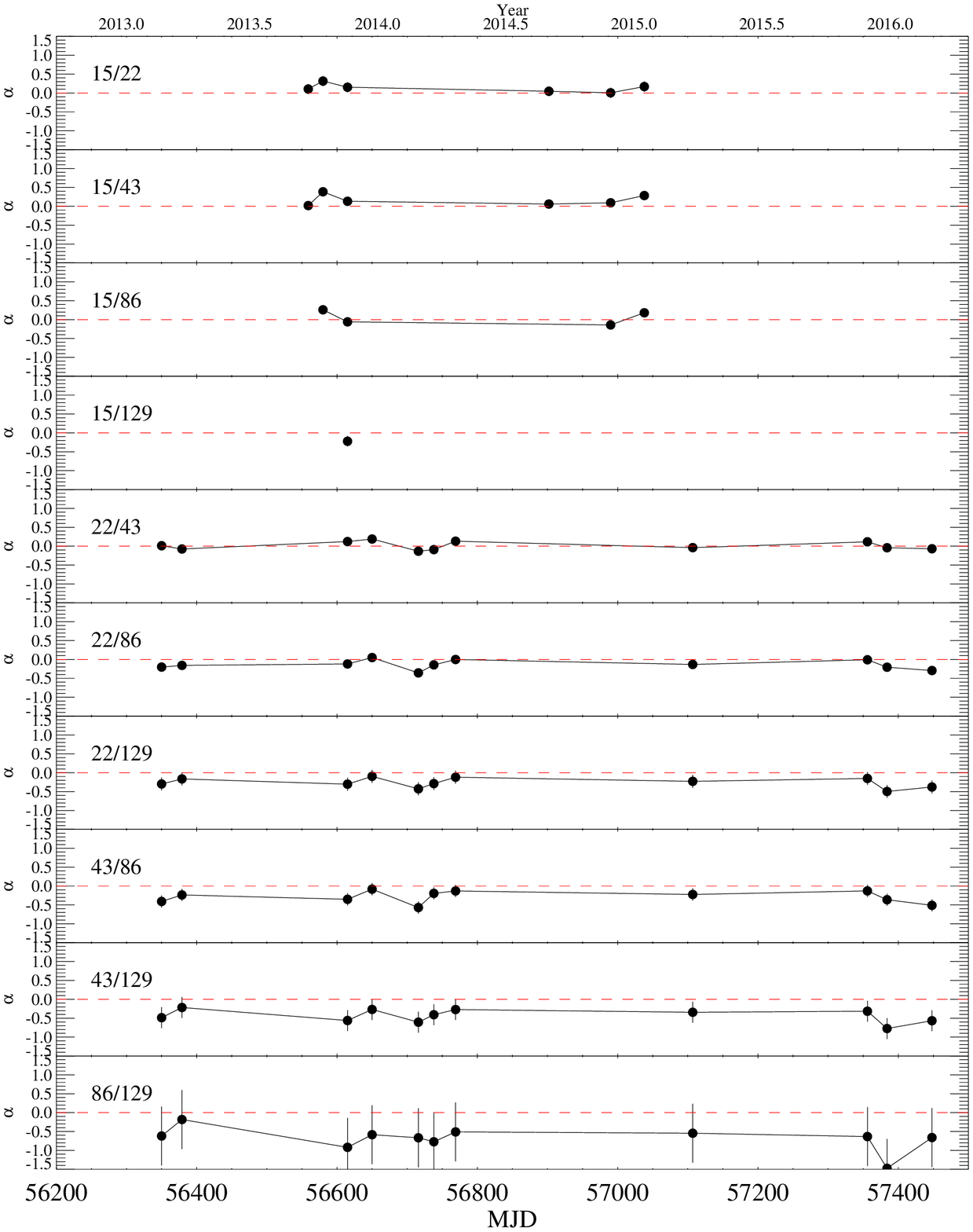}
\caption{Multiwavelengths spectral indices of S5 0716+714.
\label{fig-multi-sindex}}
\end{figure*}
\begin{figure*}[!t]
\epsscale{1.9}
\plotone{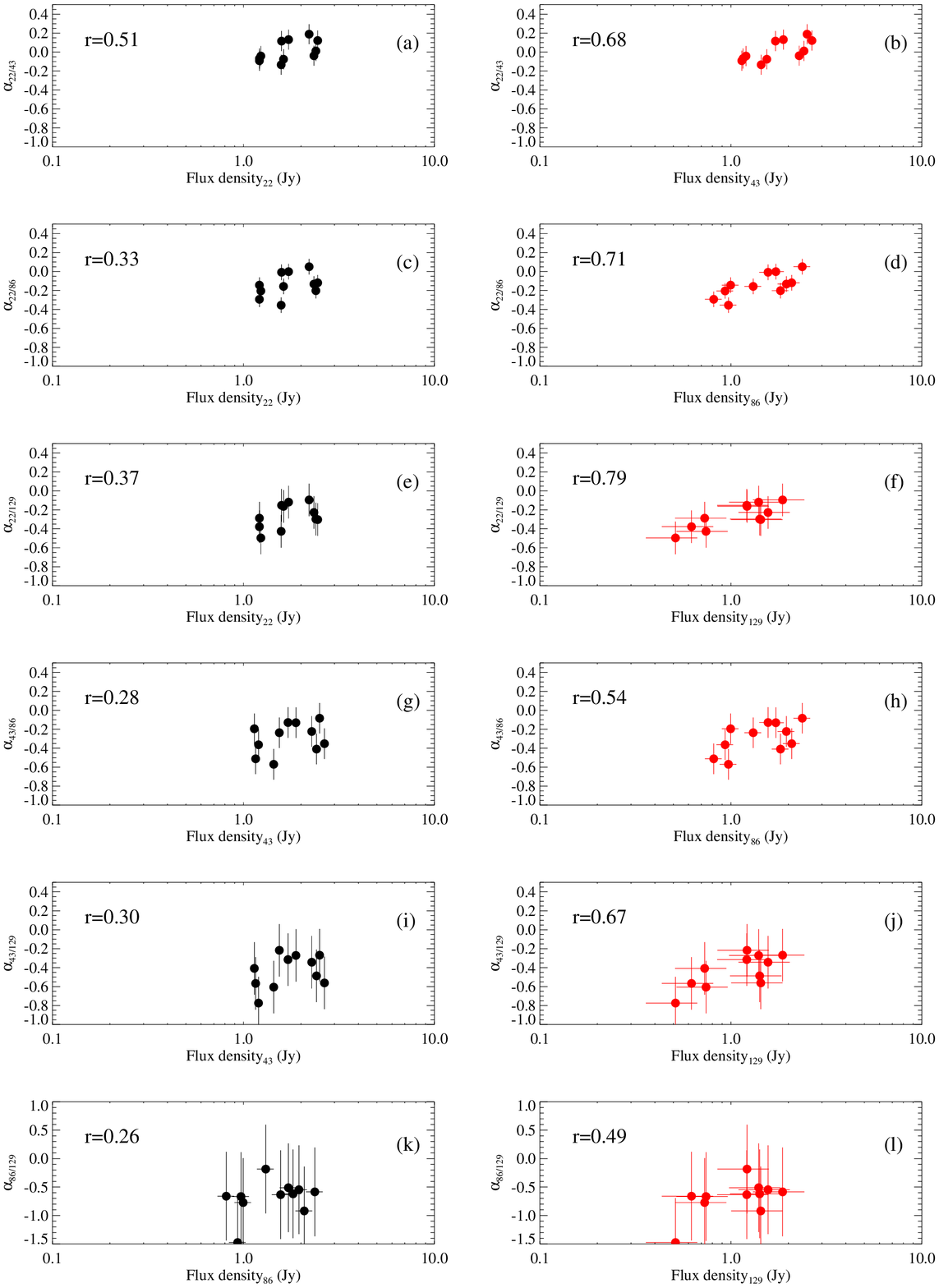}
\caption{Spectral index at lower frequency (left panel) and at higher frequency (right panel) as a function of flux density for each pair of frequencies:
(a)-(b) for 22-43~GHz,
(c)-(d) for 22-86~GHz,
(e)-(f) for 22-129~GHz,
(g)-(h) for 43-86~GHz,
(i)-(j) for 43-129~GHz,
and (k)-(l) for 86-129~GHz.
The Pearson coefficient for a potential correlation is shown at each panel at the upper, left corner. 
\label{fig-flux-sindex}}
\end{figure*}
\begin{figure*}[!t]
\epsscale{1.9}
\plotone{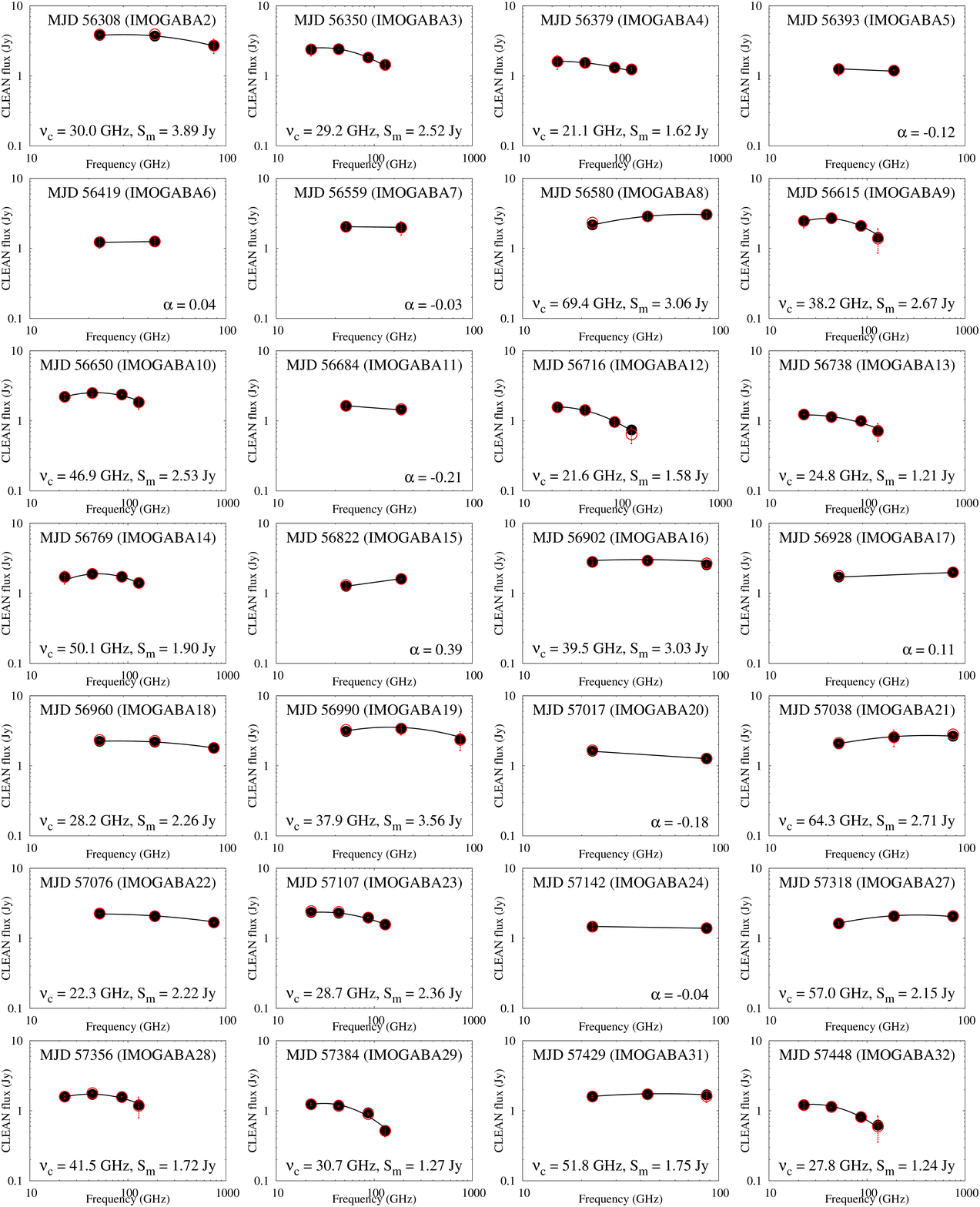}
\caption{Spectra of CLEAN flux density (black dot)
	and core flux density (red circle) at 22-129~GHz for 28 epochs.
	Black solid lines are the best fitting power law.
\label{fig-spind}}
\end{figure*}
\begin{figure*}[!t]
\epsscale{2.3}
\plotone{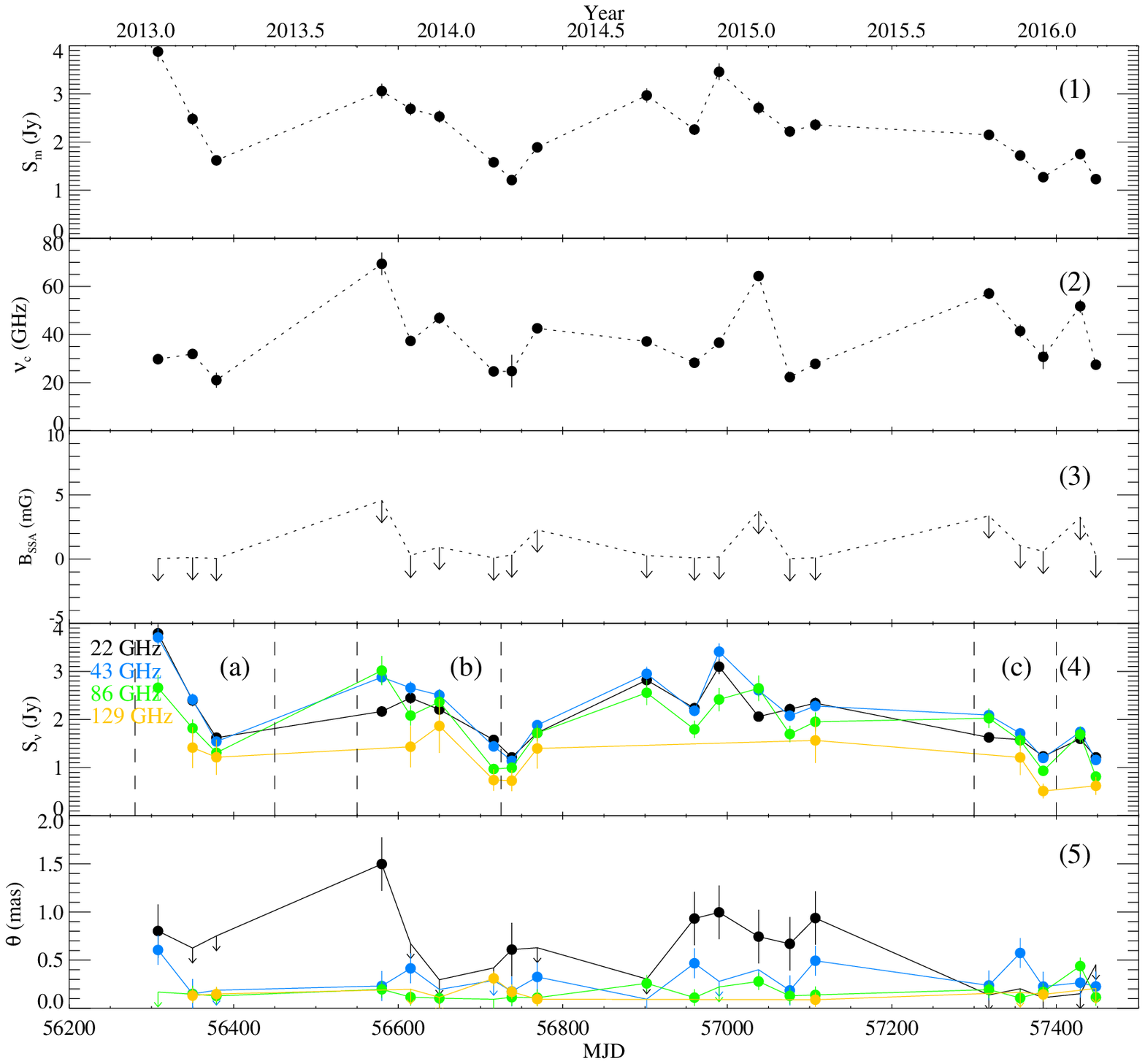}
\caption{Variation of the peak flux densities (1), turnover frequencies (2), magnetic fields (3), KVN flux densities (4), and deconvolved core sizes (5) in the time domain.
\label{fig-multi-curve}}
\end{figure*}
\begin{figure*}[!t]
\epsscale{1.0}
\plotone{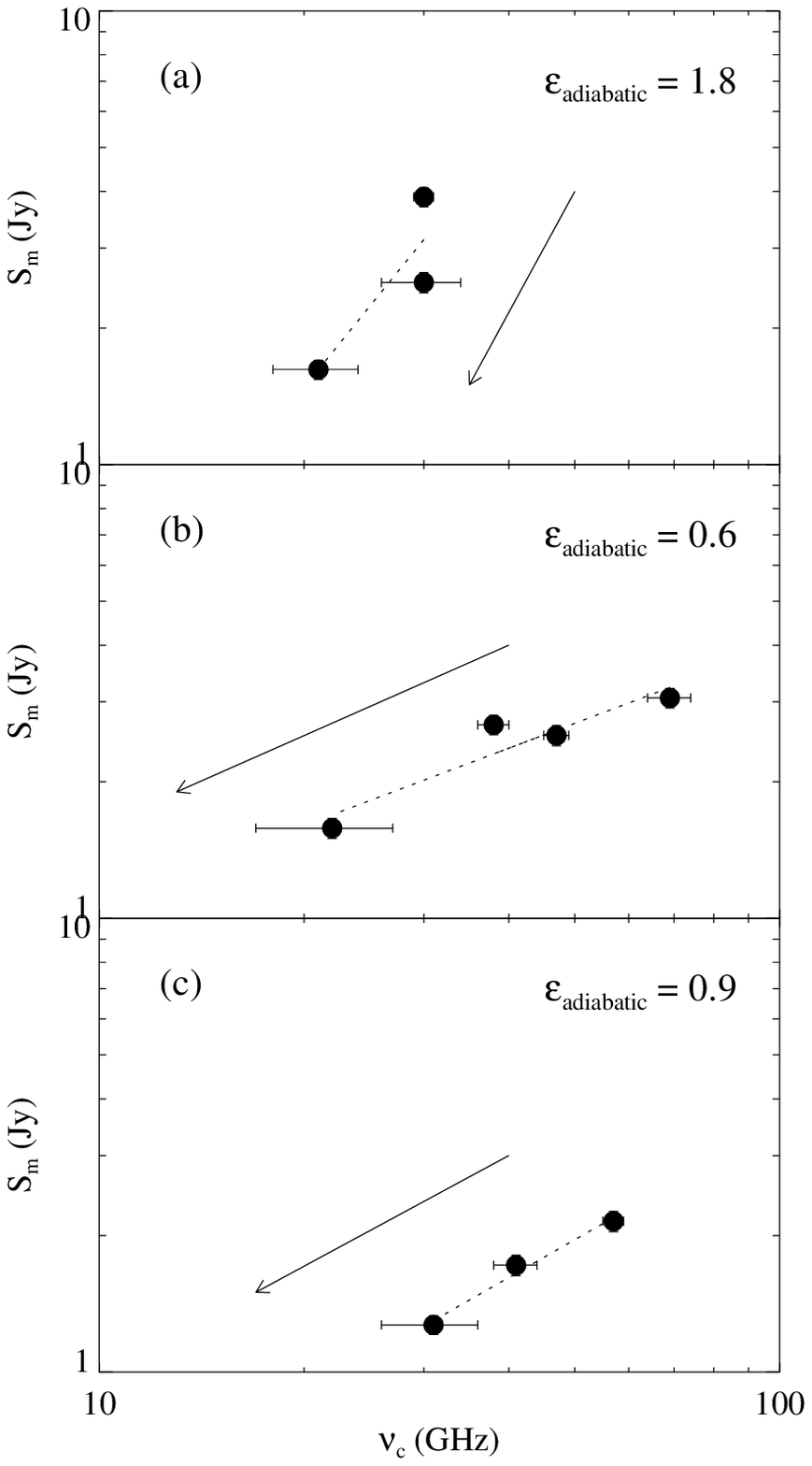}
\caption{Evolution of the peak flux density $S_{\rm m}$ and the turnover frequency $\nu_{\rm c}$ for periods (a), (b), and (c). The arrows show the time flow in each period.
\label{fig-nuc-sm}}
\end{figure*}
\begin{figure*}[!t]
\epsscale{1.9}
\plotone{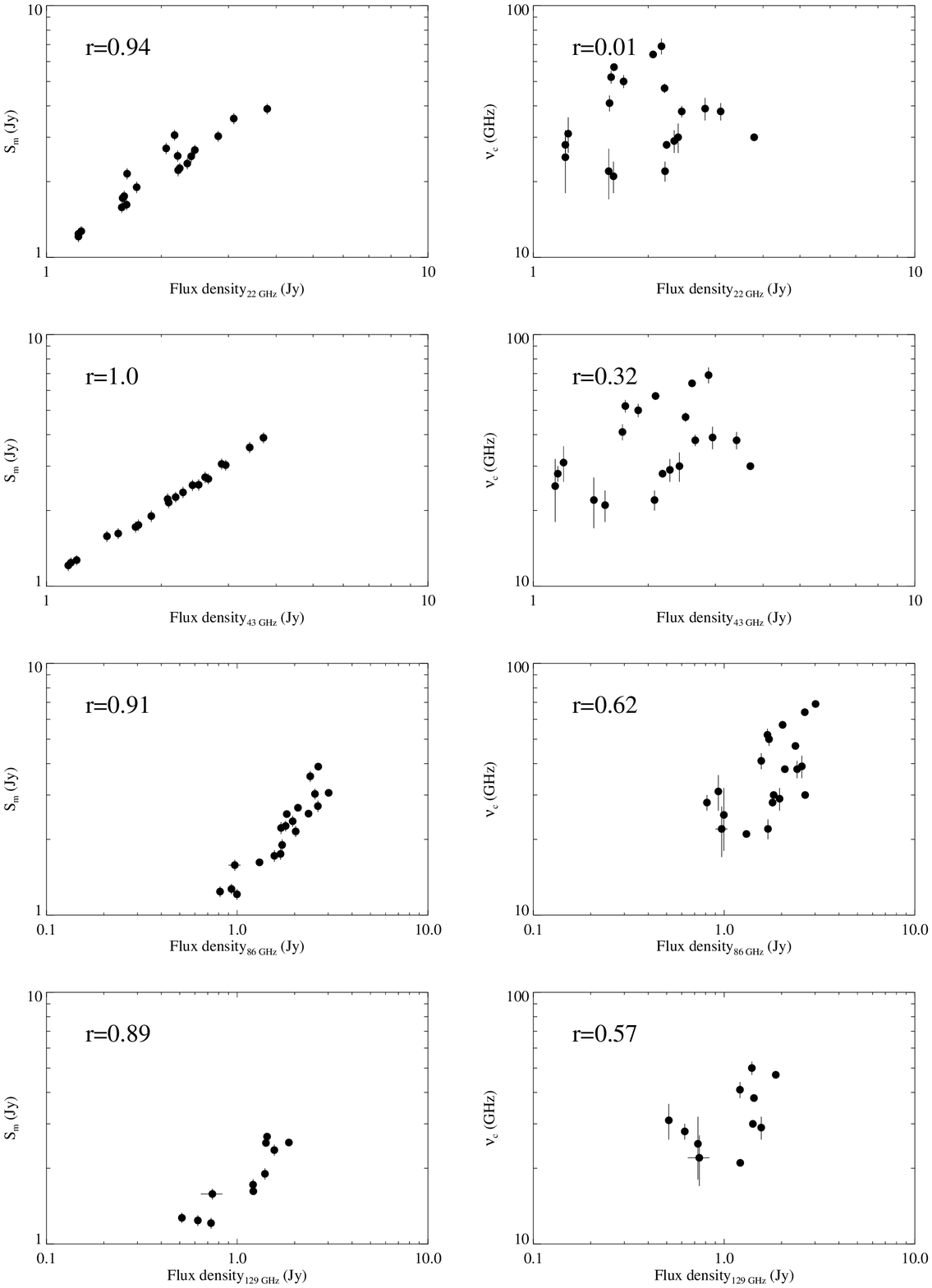}
\caption{Peak flux density (left panels) and turnover frequency (right panels) as a function of the flux density for the different observing frequencies. The Pearson coefficient for a potential correlation is shown at each panel at the upper, left corner.
\label{fig-flux-sm_nuc}}
\end{figure*}



\clearpage

\begin{deluxetable}{llcccrrrrrr}
\tabletypesize{\scriptsize}
\tablecaption{Image parameters\label{t2}}
\tablewidth{0pt}
\tablehead{
\colhead{Epoch} & 
\colhead{MJD} & 
\colhead{Band} & 
\colhead{$B_{\rm maj}$} &
\colhead{$B_{\rm min}$} & 
\colhead{$B_{\rm PA}$} & 
\colhead{$S_{\rm KVN}$} &
\colhead{$S_{\rm p}$} &
\colhead{$\sigma$} &
\colhead{$D$} &
\colhead{$\xi_{\rm r}$}\\ 
\colhead{(1)} & 
\colhead{(2)} & 
\colhead{(3)} &
\colhead{(4)} & 
\colhead{(5)} & 
\colhead{(6)} &
\colhead{(7)} &
\colhead{(8)} &
\colhead{(9)} &
\colhead{(10)} &
\colhead{(11)}
}
\startdata
2013 Jan 16&     56308 &K   &      5.471 &      4.052 &  $-$86.4 &       3.791 &       3.787 &          16 &    234 &  0.61 \\
           &           &Q   &      2.723 &      2.030 &  $-$85.7 &       3.705 &       3.703 &          29 &    129 &  0.62 \\
           &           &W   &      1.354 &      0.914 &       89.3 &       2.660 &       2.694 &          74 &     36 &  0.41 \\
2013 Feb 27&     56350 &K   &      5.509 &      4.173 &  $-$72.8 &       2.396 &       2.409 &          10 &    251 &  0.58 \\
           &           &Q   &      2.692 &      2.055 &  $-$76.5 &       2.415 &       2.399 &          28 &     86 &  0.59 \\
           &           &W   &      1.317 &      1.044 &       89.4 &       1.819 &       1.804 &          66 &     27 &  0.57 \\
           &           &D   &      0.920 &      0.701 &  $-$56.7 &       1.415 &       1.418 &          63 &     23 &  0.59 \\
2013 Mar 28&     56379 &K   &      5.373 &      4.585 &  $-$64.7 &       1.622 &       1.623 &           5 &    359 &  0.67 \\
           &           &Q   &      2.735 &      2.319 &  $-$61.0 &       1.542 &       1.543 &          11 &    142 &  0.69 \\
           &           &W   &      1.308 &      1.093 &  $-$77.2 &       1.309 &       1.298 &          25 &     51 &  0.59 \\
           &           &D   &      0.889 &      0.610 &  $-$79.5 &       1.215 &       1.191 &          57 &     21 &  0.56 \\
2013 Apr 11&     56393 &K   &      5.330 &      4.294 &  $-$79.0 &       1.262 &       1.261 &           6 &    200 &  0.59 \\
           &           &Q   &      2.703 &      2.055 &  $-$80.2 &       1.170 &       1.173 &          20 &     59 &  0.60 \\
2013 May 07&     56419 &K   &      6.443 &      4.533 &  $-$48.4 &       1.219 &       1.237 &           7 &    173 &  0.54 \\
           &           &Q   &      3.291 &      2.241 &  $-$55.5 &       1.252 &       1.244 &          21 &     59 &  0.55 \\
2013 Sep 24&     56559 &K   &      5.514 &      4.145 &  $-$62.6 &       2.041 &       2.041 &          13 &    153 &  0.65 \\
           &           &Q   &      2.744 &      2.154 &  $-$57.7 &       2.000 &       2.007 &          27 &     74 &  0.54 \\
2013 Oct 15&     56580 &K   &      5.663 &      4.634 &  $-$73.4 &       2.167 &       2.160 &          22 &     96 &  0.56 \\
           &           &Q   &      3.414 &      2.439 &  $-$45.2 &       2.879 &       2.864 &          41 &     70 &  0.49 \\
           &           &W   &      1.806 &      1.228 &  $-$29.7 &       3.016 &       3.006 &         126 &     24 &  0.47 \\
2013 Nov 19&     56615 &K   &      5.849 &      4.196 &  $-$61.1 &       2.449 &       2.486 &          20 &    122 &  0.61 \\
           &           &Q   &      2.857 &      2.005 &  $-$70.4 &       2.657 &       2.634 &          20 &    131 &  0.62 \\
           &           &W   &      1.440 &      0.969 &  $-$73.9 &       2.082 &       2.077 &          28 &     73 &  0.53 \\
           &           &D   &      0.934 &      0.645 &  $-$63.0 &       1.434 &       1.427 &          31 &     46 &  0.57 \\
2013 Dec 24&     56650 &K   &      5.741 &      4.258 &  $-$54.1 &       2.208 &       2.208 &          10 &    220 &  0.58 \\
           &           &Q   &      2.861 &      2.151 &  $-$57.3 &       2.506 &       2.507 &          13 &    190 &  0.62 \\
           &           &W   &      1.469 &      1.042 &  $-$51.3 &       2.365 &       2.357 &          23 &    103 &  0.55 \\
           &           &D   &      1.000 &      0.663 &  $-$51.9 &       1.866 &       1.868 &          29 &     64 &  0.58 \\
2014 Jan 27&     56684 &K   &      5.880 &      4.459 &  $-$53.0 &       1.642 &       1.643 &          12 &    142 &  0.63 \\
           &           &Q   &      2.890 &      2.144 &  $-$57.1 &       1.433 &       1.453 &          22 &     65 &  0.54 \\
2014 Feb 28&     56716 &K   &      6.768 &      3.519 &  $-$51.7 &       1.576 &       1.574 &          19 &     84 &  0.57 \\
           &           &Q   &      3.341 &      1.861 &  $-$46.4 &       1.441 &       1.434 &          27 &     54 &  0.57 \\
           &           &W   &      2.085 &      0.841 &  $-$43.8 &       0.971 &       0.968 &          69 &     14 &  0.57 \\
           &           &D   &      1.372 &      0.611 &  $-$36.3 &       0.741 &       0.604 &          96 &      6 &  0.45 \\
2014 Mar 22&     56738 &K   &      5.273 &      3.800 &  $-$68.0 &       1.213 &       1.213 &           6 &    217 &  0.55 \\
           &           &Q   &      2.589 &      1.959 &  $-$68.2 &       1.141 &       1.129 &          12 &     92 &  0.61 \\
           &           &W   &      1.407 &      0.912 &  $-$61.8 &       0.997 &       0.987 &          22 &     45 &  0.58 \\
           &           &D   &      0.869 &      0.642 &  $-$74.3 &       0.729 &       0.712 &          24 &     29 &  0.68 \\
2014 Apr 22&     56769 &K   &      5.274 &      3.775 &  $-$71.7 &       1.724 &       1.733 &           7 &    257 &  0.68 \\
           &           &Q   &      2.641 &      1.887 &  $-$69.1 &       1.883 &       1.859 &          18 &    106 &  0.69 \\
           &           &W   &      1.390 &      0.884 &  $-$79.4 &       1.720 &       1.706 &          32 &     54 &  0.63 \\
           &           &D   &      0.952 &      0.580 &  $-$79.1 &       1.398 &       1.383 &          28 &     49 &  0.54 \\
2014 Jun 13&     56822 &K   &      5.330 &      4.294 &  $-$79.0 &       1.262 &       1.261 &           6 &    200 &  0.59 \\
           &           &Q   &      2.645 &      1.943 &  $-$68.8 &       1.621 &       1.585 &          33 &     49 &  0.62 \\
2014 Sep 01&     56902 &K   &      5.346 &      3.692 &  $-$87.3 &       2.820 &       2.815 &          13 &    215 &  0.69 \\
           &           &Q   &      2.753 &      1.792 &  $-$83.1 &       2.950 &       2.939 &          20 &    145 &  0.56 \\
           &           &W   &      1.378 &      0.898 &  $-$88.8 &       2.556 &       2.550 &          31 &     83 &  0.55 \\
2014 Sep 27&     56928 &K   &      5.723 &      3.474 &  $-$67.0 &       1.711 &       1.709 &           7 &    232 &  0.68 \\
           &           &W   &      1.544 &      0.805 &  $-$72.1 &       1.984 &       1.965 &          61 &     32 &  0.59 \\
2014 Oct 29&     56960 &K   &      5.862 &      3.422 &  $-$74.4 &       2.235 &       2.233 &          12 &    190 &  0.66 \\
           &           &Q   &      2.904 &      1.739 &  $-$72.2 &       2.181 &       2.172 &          22 &     98 &  0.62 \\
           &           &W   &      1.636 &      0.799 &  $-$70.0 &       1.795 &       1.779 &          33 &     53 &  0.57 \\
2014 Nov 28&     56990 &K   &      5.657 &      3.436 &  $-$71.8 &       3.097 &       3.089 &          24 &    128 &  0.63 \\
           &           &Q   &      2.761 &      1.770 &  $-$73.8 &       3.410 &       3.428 &          19 &    183 &  0.65 \\
           &           &W   &      1.467 &      0.834 &  $-$72.9 &       2.415 &       2.406 &          55 &     44 &  0.69 \\
2014 Dec 25&     57017 &K   &      5.431 &      3.672 &  $-$72.3 &       1.613 &       1.610 &           9 &    183 &  0.67 \\
           &           &W   &      1.368 &      0.912 &  $-$69.2 &       1.265 &       1.249 &          28 &     45 &  0.60 \\
2015 Jan 15&     57038 &K   &      6.055 &      3.358 &  $-$70.9 &       2.060 &       2.058 &           8 &    265 &  0.63 \\
           &           &Q   &      2.881 &      1.752 &  $-$72.4 &       2.606 &       2.603 &          11 &    228 &  0.70 \\
           &           &W   &      1.555 &      0.825 &  $-$70.3 &       2.648 &       2.636 &          28 &     93 &  0.65 \\
2015 Feb 23&     57076 &K   &      5.360 &      3.747 &  $-$71.2 &       2.214 &       2.214 &           5 &    463 &  0.70 \\
           &           &Q   &      2.604 &      1.961 &  $-$72.7 &       2.078 &       2.051 &          20 &    105 &  0.57 \\
           &           &W   &      1.471 &      0.971 &  $-$53.6 &       1.698 &       1.665 &          72 &     23 &  0.49 \\
2015 Mar 26&     57107 &K   &      5.400 &      3.626 &  $-$80.5 &       2.340 &       2.336 &           7 &    353 &  0.60 \\
           &           &Q   &      2.599 &      1.907 &  $-$76.5 &       2.280 &       2.272 &           7 &    339 &  0.62 \\
           &           &W   &      1.364 &      0.899 &  $-$80.9 &       1.953 &       1.939 &          29 &     67 &  0.67 \\
           &           &D   &      0.900 &      0.621 &  $-$75.8 &       1.565 &       1.550 &          41 &     38 &  0.54 \\
2015 Apr 30&     57142 &K   &      5.452 &      3.729 &  $-$71.8 &       1.476 &       1.475 &          13 &    118 &  0.71 \\
           &           &W   &      1.440 &      0.862 &  $-$73.5 &       1.390 &       1.373 &          26 &     53 &  0.67 \\
2015 Oct 23&     57318 &K   &      5.647 &      3.843 &  $-$66.8 &       1.627 &       1.626 &          14 &    112 &  0.63 \\
           &           &Q   &      2.950 &      1.895 &  $-$58.9 &       2.091 &       2.049 &          41 &     50 &  0.53 \\
           &           &W   &      1.636 &      0.912 &  $-$53.1 &       2.026 &       2.021 &          82 &     25 &  0.49 \\
2015 Nov 30&     57356 &K   &      5.398 &      3.773 &  $-$71.8 &       1.584 &       1.585 &           9 &    169 &  0.58 \\
           &           &Q   &      2.738 &      1.841 &  $-$67.4 &       1.712 &       1.679 &          10 &    175 &  0.53 \\
           &           &W   &      1.451 &      0.883 &  $-$68.2 &       1.566 &       1.555 &          18 &     88 &  0.64 \\
           &           &D   &      0.956 &      0.598 &  $-$68.5 &       1.211 &       1.206 &          24 &     49 &  0.55 \\
2015 Dec 28&     57384 &K   &      5.327 &      3.781 &  $-$68.2 &       1.234 &       1.234 &           8 &    155 &  0.62 \\
           &           &Q   &      2.647 &      1.950 &  $-$68.5 &       1.200 &       1.161 &           6 &    204 &  0.79 \\
           &           &W   &      1.415 &      0.906 &  $-$68.6 &       0.933 &       0.880 &          12 &     71 &  0.66 \\
           &           &D   &      0.922 &      0.625 &  $-$64.5 &       0.513 &       0.508 &          10 &     50 &  0.71 \\
2016 Jan 13&     57400 &K   &      5.310 &      3.952 &  $-$66.1 &       1.261 &       1.260 &           5 &    240 &  0.74 \\
2016 Feb 11&     57429 &K   &      5.548 &      4.153 &  $-$82.2 &       1.599 &       1.598 &           9 &    177 &  0.63 \\
           &           &Q   &      2.863 &      2.086 &  $-$73.9 &       1.742 &       1.698 &          17 &    103 &  0.69 \\
           &           &W   &      1.557 &      0.990 &  $-$73.5 &       1.688 &       1.482 &          44 &     34 &  0.65 \\
2016 Mar 01&     57448 &K   &      5.250 &      3.856 &  $-$66.6 &       1.213 &       1.213 &           5 &    222 &  0.63 \\
           &           &Q   &      2.609 &      1.964 &  $-$66.9 &       1.159 &       1.125 &           6 &    193 &  0.65 \\
           &           &W   &      1.350 &      0.935 &  $-$68.5 &       0.813 &       0.805 &           9 &     92 &  0.70 \\
           &           &D   &      0.868 &      0.669 &  $-$57.5 &       0.622 &       0.620 &           9 &     67 &  0.60 \\
\enddata
\tablecomments{ 
Column designation: 1~-~Date; 2~-~modified Julian date; 3~-~observing frequency band: 
K~-~22~GHz band; Q~-~43~GHz band; W~-~86~GHz band; D~-~129~GHz band; 
4-6~-~restoring beam:
4~-~major axis [mas];
5~-~minor axis [mas];
6~-~position angle of the major axis [degree];
7~-~total cleaned KVN flux density [Jy];
8~-~peak flux density [Jy beam$^{-1}$];
9~-~off-source RMS in the image [Jy beam$^{-1}$];
10~-~dynamic range of the image;
11~-~quality of the residual noise in the image (i.e., ratio of the image root-mean-square noise to its mathematical expectation).
}
\end{deluxetable}

\begin{deluxetable}{llcccrrrr}
\tabletypesize{\scriptsize}
\tablecaption{Modelfit parameters\label{t3}}
\tablewidth{0pt}
\tablehead{
\colhead{Epoch} & 
\colhead{MJD} & 
\colhead{Band} & 
\colhead{$S_{\rm tot}$} &
\colhead{$S_{\rm peak}$} & 
\colhead{$d$} & 
\colhead{$r$} &
\colhead{$\theta$} &
\colhead{$T_{\rm b}$} \\ 
\colhead{(1)} & 
\colhead{(2)} & 
\colhead{(3)} &
\colhead{(4)} & 
\colhead{(5)} & 
\colhead{(6)} &
\colhead{(7)} &
\colhead{(8)} &
\colhead{(9)}
}
\startdata
2013 Jan 16  &       56308   &  K   &    3.89$\pm$ 0.04 &    3.79$\pm$ 0.03 &   0.80$\pm$ 0.01    &              ... &             ... &    1.12$\pm$  0.02 \\
             &               &  Q   &    3.93$\pm$ 0.06 &    3.70$\pm$ 0.04 &   0.60$\pm$ 0.01    &              ... &             ... &    1.99$\pm$  0.04 \\
             &               &  W   &    2.72$\pm$ 0.63 &    2.62$\pm$ 0.44 &  $<${\it  0.17}     &              ... &             ... &  $>${\it  17.74} \\
2013 Feb 27  &       56350   &  K   &    2.39$\pm$ 0.47 &    2.41$\pm$ 0.33 &  $<${\it  0.63}     &              ... &             ... &  $>${\it   1.13} \\
             &               &  Q   &    2.41$\pm$ 0.02 &    2.40$\pm$ 0.02 &   0.15$\pm$ 0.00    &              ... &             ... &   20.08$\pm$  0.27 \\
             &               &  W   &    1.83$\pm$ 0.05 &    1.80$\pm$ 0.03 &   0.14$\pm$ 0.00    &              ... &             ... &   16.11$\pm$  0.62 \\
             &               &  D   &    1.44$\pm$ 0.16 &    1.42$\pm$ 0.12 &   0.13$\pm$ 0.01    &              ... &             ... &   15.57$\pm$  2.53 \\
2013 Mar 28  &       56379   &  K   &    1.60$\pm$ 0.37 &    1.62$\pm$ 0.26 &  $<${\it  0.75}     &              ... &             ... &  $>${\it   0.53} \\
             &               &  Q   &    1.54$\pm$ 0.17 &    1.54$\pm$ 0.12 &  $<${\it  0.19}     &              ... &             ... &  $>${\it   8.17} \\
             &               &  W   &    1.31$\pm$ 0.02 &    1.30$\pm$ 0.02 &   0.13$\pm$ 0.00    &              ... &             ... &   15.05$\pm$  0.38 \\
             &               &  D   &    1.24$\pm$ 0.06 &    1.19$\pm$ 0.04 &   0.15$\pm$ 0.01    &              ... &             ... &   10.61$\pm$  0.79 \\
2013 Apr 11  &       56393   &  K   &    1.25$\pm$ 0.25 &    1.26$\pm$ 0.17 &  $<${\it  0.63}     &              ... &             ... &  $>${\it   0.59} \\
             &               &  Q   &    1.19$\pm$ 0.02 &    1.17$\pm$ 0.01 &   0.31$\pm$ 0.00    &              ... &             ... &    2.28$\pm$  0.05 \\
2013 May 07  &       56419   &  K   &    1.23$\pm$ 0.22 &    1.24$\pm$ 0.16 &  $<${\it  0.66}     &              ... &             ... &  $>${\it   0.53} \\
             &               &  Q   &    1.25$\pm$ 0.16 &    1.24$\pm$ 0.11 &  $<${\it  0.22}     &              ... &             ... &  $>${\it   4.68} \\
2013 Sep 24  &       56559   &  K   &    2.04$\pm$ 0.21 &    2.04$\pm$ 0.15 &  $<${\it  0.33}     &              ... &             ... &  $>${\it   3.45} \\
             &               &  Q   &    1.98$\pm$ 0.44 &    2.01$\pm$ 0.32 &  $<${\it  0.36}     &              ... &             ... &  $>${\it   2.81} \\
2013 Oct 15  &       56580   &  K   &    2.33$\pm$ 0.06 &    2.16$\pm$ 0.04 &   1.50$\pm$ 0.03    &              ... &             ... &    0.19$\pm$  0.01 \\
             &               &  Q   &    2.87$\pm$ 0.31 &    2.87$\pm$ 0.22 &   0.23$\pm$ 0.02    &              ... &             ... &    9.94$\pm$  1.52 \\
             &               &  W   &    3.04$\pm$ 0.30 &    3.01$\pm$ 0.21 &   0.20$\pm$ 0.01    &              ... &             ... &   14.65$\pm$  2.05 \\
2013 Nov 19  &       56615   &  K   &    2.46$\pm$ 0.50 &    2.49$\pm$ 0.36 &  $<${\it  0.67}     &              ... &             ... &  $>${\it   1.01} \\
             &               &  Q   &    2.70$\pm$ 0.01 &    2.63$\pm$ 0.01 &   0.41$\pm$ 0.00    &              ... &             ... &    2.92$\pm$  0.02 \\
             &               &  W   &    2.10$\pm$ 0.03 &    2.08$\pm$ 0.02 &   0.12$\pm$ 0.00    &              ... &             ... &   29.33$\pm$  0.63 \\
             &               &  D   &    1.38$\pm$ 0.53 &    1.43$\pm$ 0.38 &  $<${\it  0.20}     &              ... &             ... &  $>${\it   6.63} \\
2013 Dec 24  &       56650   &  K   &    2.20$\pm$ 0.20 &    2.21$\pm$ 0.14 &  $<${\it  0.30}     &              ... &             ... &  $>${\it   4.65} \\
             &               &  Q   &    2.50$\pm$ 0.30 &    2.51$\pm$ 0.21 &  $<${\it  0.20}     &              ... &             ... &  $>${\it  12.11} \\
             &               &  W   &    2.37$\pm$ 0.02 &    2.36$\pm$ 0.02 &   0.10$\pm$ 0.00    &              ... &             ... &   40.61$\pm$  0.53 \\
             &               &  D   &    1.84$\pm$ 0.39 &    1.87$\pm$ 0.27 &  $<${\it  0.11}     &              ... &             ... &  $>${\it  26.68} \\
2014 Jan 27  &       56684   &  K   &    1.64$\pm$ 0.15 &    1.64$\pm$ 0.10 &  $<${\it  0.30}     &              ... &             ... &  $>${\it   3.32} \\
             &               &  Q   &    1.47$\pm$ 0.02 &    1.45$\pm$ 0.02 &   0.28$\pm$ 0.00    &              ... &             ... &    3.40$\pm$  0.07 \\
2014 Feb 28  &       56716   &  K   &    1.57$\pm$ 0.20 &    1.57$\pm$ 0.14 &  $<${\it  0.42}     &              ... &             ... &  $>${\it   1.66} \\
             &               &  Q   &    1.42$\pm$ 0.25 &    1.44$\pm$ 0.18 &  $<${\it  0.29}     &              ... &             ... &  $>${\it   3.17} \\
             &               &  W   &    0.96$\pm$ 0.10 &    0.97$\pm$ 0.07 &  $<${\it  0.09}     &              ... &             ... &  $>${\it  20.69} \\
             &               &  D   &    0.64$\pm$ 0.17 &    0.58$\pm$ 0.11 &   0.31$\pm$ 0.06    &              ... &             ... &    1.24$\pm$  0.49 \\
2014 Mar 22  &       56738   &  K   &    1.23$\pm$ 0.01 &    1.21$\pm$ 0.01 &   0.61$\pm$ 0.00    &              ... &             ... &    0.62$\pm$  0.01 \\
             &               &  Q   &    1.14$\pm$ 0.01 &    1.13$\pm$ 0.01 &   0.17$\pm$ 0.00    &              ... &             ... &    6.94$\pm$  0.12 \\
             &               &  W   &    1.00$\pm$ 0.02 &    0.99$\pm$ 0.01 &   0.11$\pm$ 0.00    &              ... &             ... &   14.20$\pm$  0.39 \\
             &               &  D   &    0.71$\pm$ 0.20 &    0.71$\pm$ 0.14 &   0.17$\pm$ 0.03    &              ... &             ... &    4.56$\pm$  1.86 \\
2014 Apr 22  &       56769   &  K   &    1.71$\pm$ 0.36 &    1.73$\pm$ 0.26 &  $<${\it  0.63}     &              ... &             ... &  $>${\it   0.80} \\
             &               &  Q   &    1.89$\pm$ 0.04 &    1.86$\pm$ 0.03 &   0.33$\pm$ 0.00    &              ... &             ... &    3.32$\pm$  0.09 \\
             &               &  W   &    1.72$\pm$ 0.03 &    1.71$\pm$ 0.02 &   0.11$\pm$ 0.00    &              ... &             ... &   26.84$\pm$  0.64 \\
             &               &  D   &    1.41$\pm$ 0.03 &    1.38$\pm$ 0.02 &   0.09$\pm$ 0.00    &              ... &             ... &   29.46$\pm$  1.02 \\
2014 Jun 13  &       56822   &  K   &    1.31$\pm$ 0.14 &    1.26$\pm$ 0.10 &   0.92$\pm$ 0.07    &              ... &             ... &    0.29$\pm$  0.05 \\
             &               &  Q   &    1.60$\pm$ 0.03 &    1.58$\pm$ 0.02 &   0.27$\pm$ 0.00    &              ... &             ... &    3.96$\pm$  0.11 \\
2014 Sep 01  &       56902   &  K   &    2.81$\pm$ 0.29 &    2.81$\pm$ 0.20 &  $<${\it  0.30}     &              ... &             ... &  $>${\it   5.63} \\
             &               &  Q   &    2.94$\pm$ 0.19 &    2.94$\pm$ 0.13 &  $<${\it  0.09}     &              ... &             ... &  $>${\it  61.14} \\
             &               &  W   &    2.68$\pm$ 0.21 &    2.55$\pm$ 0.14 &   0.26$\pm$ 0.01    &              ... &             ... &    7.44$\pm$  0.83 \\
2014 Sep 27  &       56928   &  K   &    1.78$\pm$ 0.02 &    1.71$\pm$ 0.01 &   0.90$\pm$ 0.01    &              ... &             ... &    0.41$\pm$  0.01 \\
             &               &  W   &    1.99$\pm$ 0.05 &    1.97$\pm$ 0.03 &   0.12$\pm$ 0.00    &              ... &             ... &   23.62$\pm$  0.77 \\
2014 Oct 29  &       56960   &  K   &    2.33$\pm$ 0.03 &    2.23$\pm$ 0.02 &   0.93$\pm$ 0.01    &              ... &             ... &    0.50$\pm$  0.01 \\
             &               &  Q   &    2.27$\pm$ 0.03 &    2.17$\pm$ 0.02 &   0.47$\pm$ 0.01    &              ... &             ... &    1.92$\pm$  0.04 \\
             &               &  W   &    1.80$\pm$ 0.03 &    1.78$\pm$ 0.02 &   0.11$\pm$ 0.00    &              ... &             ... &   27.51$\pm$  0.75 \\
2014 Nov 28  &       56990   &  K   &    3.25$\pm$ 0.09 &    3.09$\pm$ 0.06 &   1.00$\pm$ 0.02    &              ... &             ... &    0.61$\pm$  0.02 \\
             &               &  Q   &    3.40$\pm$ 0.65 &    3.43$\pm$ 0.46 &  $<${\it  0.28}     &              ... &             ... &  $>${\it   8.08} \\
             &               &  W   &    2.35$\pm$ 0.71 &    2.41$\pm$ 0.51 &  $<${\it  0.22}     &              ... &             ... &  $>${\it   8.93} \\
2014 Dec 25  &       57017   &  K   &    1.66$\pm$ 0.03 &    1.61$\pm$ 0.02 &   0.83$\pm$ 0.01    &              ... &             ... &    0.45$\pm$  0.01 \\
             &               &  W   &    1.27$\pm$ 0.04 &    1.25$\pm$ 0.03 &   0.14$\pm$ 0.00    &              ... &             ... &   11.18$\pm$  0.48 \\
2015 Jan 15  &       57038   &  K   &    2.12$\pm$ 0.02 &    2.06$\pm$ 0.02 &   0.74$\pm$ 0.01    &              ... &             ... &    0.71$\pm$  0.01 \\
             &               &  Q   &    2.56$\pm$ 0.69 &    2.60$\pm$ 0.49 &  $<${\it  0.40}     &              ... &             ... &  $>${\it   2.96} \\
             &               &  W   &    2.81$\pm$ 0.06 &    2.64$\pm$ 0.04 &   0.28$\pm$ 0.00    &              ... &             ... &    6.73$\pm$  0.21 \\
2015 Feb 23  &       57076   &  K   &    2.26$\pm$ 0.02 &    2.21$\pm$ 0.01 &   0.67$\pm$ 0.00    &              ... &             ... &    0.93$\pm$  0.01 \\
             &               &  Q   &    2.06$\pm$ 0.02 &    2.05$\pm$ 0.01 &   0.18$\pm$ 0.00    &              ... &             ... &   11.17$\pm$  0.14 \\
             &               &  W   &    1.68$\pm$ 0.04 &    1.67$\pm$ 0.03 &   0.13$\pm$ 0.00    &              ... &             ... &   19.03$\pm$  0.63 \\
2015 Mar 26  &       57107   &  K   &    2.44$\pm$ 0.02 &    2.34$\pm$ 0.01 &   0.94$\pm$ 0.00    &              ... &             ... &    0.51$\pm$  0.01 \\
             &               &  Q   &    2.38$\pm$ 0.02 &    2.27$\pm$ 0.01 &   0.49$\pm$ 0.00    &              ... &             ... &    1.81$\pm$  0.02 \\
             &               &  W   &    1.97$\pm$ 0.04 &    1.94$\pm$ 0.03 &   0.14$\pm$ 0.00    &              ... &             ... &   19.13$\pm$  0.54 \\
             &               &  D   &    1.57$\pm$ 0.03 &    1.55$\pm$ 0.02 &   0.09$\pm$ 0.00    &              ... &             ... &   37.54$\pm$  0.90 \\
2015 Apr 30  &       57142   &  K   &    1.47$\pm$ 0.22 &    1.48$\pm$ 0.16 &  $<${\it  0.45}     &              ... &             ... &  $>${\it   1.34} \\
             &               &  W   &    1.39$\pm$ 0.03 &    1.38$\pm$ 0.02 &   0.13$\pm$ 0.00    &              ... &             ... &   15.74$\pm$  0.44 \\
2015 Oct 23  &       57318   &  K   &    1.63$\pm$ 0.07 &    1.63$\pm$ 0.05 &  $<${\it  0.14}     &              ... &             ... &  $>${\it  16.01} \\
             &               &  Q   &    2.07$\pm$ 0.03 &    2.05$\pm$ 0.02 &   0.24$\pm$ 0.00    &              ... &             ... &    6.82$\pm$  0.15 \\
             &               &  W   &    2.07$\pm$ 0.05 &    2.02$\pm$ 0.04 &   0.19$\pm$ 0.00    &              ... &             ... &   10.30$\pm$  0.39 \\
2015 Nov 30  &       57356   &  K   &    1.59$\pm$ 0.11 &    1.58$\pm$ 0.08 &  $<${\it  0.20}     &              ... &             ... &  $>${\it   7.22} \\
             &               &  Q   &    1.78$\pm$ 0.05 &    1.68$\pm$ 0.04 &   0.57$\pm$ 0.01    &              ... &             ... &    1.00$\pm$  0.04 \\
             &               &  W   &    1.57$\pm$ 0.02 &    1.56$\pm$ 0.02 &   0.11$\pm$ 0.00    &              ... &             ... &   25.86$\pm$  0.57 \\
             &               &  D   &    1.18$\pm$ 0.39 &    1.21$\pm$ 0.28 &  $<${\it  0.17}     &              ... &             ... &  $>${\it   7.98} \\
2015 Dec 28  &       57384   &  K   &    1.23$\pm$ 0.05 &    1.23$\pm$ 0.03 &  $<${\it  0.11}     &              ... &             ... &  $>${\it  18.91} \\
             &               &  Q   &    1.17$\pm$ 0.02 &    1.16$\pm$ 0.01 &   0.23$\pm$ 0.00    &              ... &             ... &    4.28$\pm$  0.11 \\
             &               &  W   &    0.90$\pm$ 0.03 &    0.88$\pm$ 0.02 &   0.17$\pm$ 0.00    &              ... &             ... &    5.75$\pm$  0.28 \\
             &               &  D   &    0.52$\pm$ 0.09 &    0.51$\pm$ 0.06 &   0.14$\pm$ 0.02    &              ... &             ... &    4.68$\pm$  1.20 \\
2016 Jan 13  &       57400   &  K   &    1.25$\pm$ 0.17 &    1.26$\pm$ 0.12 &  $<${\it  0.40}     &              ... &             ... &  $>${\it   1.42} \\
2016 Feb 11  &       57429   &  K   &    1.60$\pm$ 0.07 &    1.60$\pm$ 0.05 &  $<${\it  0.15}     &              ... &             ... &  $>${\it  13.19} \\
             &               &  Q   &    1.72$\pm$ 0.03 &    1.70$\pm$ 0.02 &   0.27$\pm$ 0.00    &              ... &             ... &    4.53$\pm$  0.11 \\
             &               &  W   &    1.62$\pm$ 0.31 &    1.46$\pm$ 0.21 &   0.44$\pm$ 0.06    &              ... &             ... &    1.56$\pm$  0.44 \\
2016 Mar 01  &       57448   &  K   &    1.21$\pm$ 0.18 &    1.21$\pm$ 0.13 &  $<${\it  0.45}     &              ... &             ... &  $>${\it   1.09} \\
             &               &  Q   &    1.13$\pm$ 0.02 &    1.12$\pm$ 0.01 &   0.23$\pm$ 0.00    &              ... &             ... &    4.11$\pm$  0.09 \\
             &               &  W   &    0.81$\pm$ 0.02 &    0.81$\pm$ 0.01 &   0.11$\pm$ 0.00    &              ... &             ... &   11.79$\pm$  0.31 \\
             &               &  D   &    0.60$\pm$ 0.24 &    0.62$\pm$ 0.17 &  $<${\it  0.21}     &              ... &             ... &  $>${\it   2.62} \\
\enddata
\tablecomments{ 
Column designation: 1~-~date; 2~-~modified Julian date;
3~-~observing frequency band: 
K~-~22~GHz band; Q~-~43~GHz band; W~-~86~GHz band; D~-~129~GHz band; 
4~-~model flux density of the component [Jy];
5~-~peak brightness of an individual component measured in the image [Jy beam$^{-1}$];
6~-~size [mas], italic numbers indicate upper limits;
7~-~radius [mas];
8~-~position angle [deg];
9~-~measured brightness temperature [$10^9$~K], italic numbers indicate lower limits.
}
\end{deluxetable}
\clearpage

\begin{table}
\scriptsize
\caption{Apparent brightness temperature, Doppler factor and size of emitting region}
\label{table:tb}
\begin{tabular}{cccccccccccccccc}
\hline
\hline
\multicolumn{5}{c}{Part I} & \multicolumn{4}{c}{Part II} & \multicolumn{4}{c}{Part III} \\
\hline
$\nu$  & $\tau_{var}$ & $\Delta S$ & $T^{app}_{B}$ & $\delta_{var}$ & $\theta$ & $\tau_{var}$ & $\Delta S$ & $T^{app}_{B}$ & $\delta_{var}$ & $\theta$ & $\tau_{var}$ & $\Delta S$ & $T^{app}_{B}$ & $\delta_{var}$ & $\theta$\\
(GHz)  &    (days)   &  (Jy) &   (K)         &                &   (mas) &    (days) &  (Jy)   &   (K)         &                &   (mas) &    (days)  &  (Jy)  &   (K)         &                &   (mas) \\
\hline
15     &     100   & 2.5   & 7$\times10^{13}$&     6.9         &   0.09  &      43   & 1.8    & 3$\times10^{14}$&     12.5         &   	0.07 &\multicolumn{4}{c}{---} \\
230    &     103   & 4.5   & 5$\times10^{11}$&     0.8          &   0.01  &\multicolumn{5}{c}{---} & 130  & 4.3    & 3$\times10^{11}$&      0.6         &    0.01 \\
\hline

\end{tabular}
\end{table}
\clearpage

\begin{table}
\tabletypesize{\scriptsize}
\caption{Spectral indices}
\label{table:sindex_range}
\begin{tabular}{lcr}
\hline
\hline
Frequency  & N  & $\alpha$\\
\hline
15-22\,GHz      &  6 &    0.01...0.3 \\
15-43\,GHz      &  6 &    0.02...0.38 \\
15-86\,GHz      &  4 & $-$0.14...0.26 \\
15-129\,GHz     &  1 & $-$0.22 \\
22-43\,GHz      & 25 & $-$0.20...0.42 \\
22-86\,GHz      & 23 & $-$0.36...0.24 \\
22-129\,GHz     & 11 & $-$0.49...$-$0.10 \\
43-86\,GHz      & 20 & $-$0.57...0.07 \\
43-129\,GHz     & 11 & $-$0.77...$-$0.22 \\
86-129\,GHz     & 11 & $-$1.5...$-$0.18 \\
\hline
\end{tabular}
\end{table}

\begin{table}
\tabletypesize{\scriptsize}
\caption{Peak flux density, turnover frequency, and B-field}
\label{table:nu_c,s_m,bfield}
\begin{tabular}{ccccccc}
\hline
\hline
MJD   & Epoch   & $S_{\rm m}$   & $\nu_{\rm c}$   & $B_{SSA}$ \\
(day) & (Year)  &   (Jy)        &  (GHz)          &  (mG)     \\
\hline
56308 & 2013.04 & 3.89$\pm$0.19 & 30$\pm$1  & $<$0.04 \\
56350 & 2013.16 & 2.52$\pm$0.13 & 30$\pm$4  & $<$0.12\\
56379 & 2013.23 & 1.62$\pm$0.08 & 21$\pm$3  & $<$0.05  \\
56580 & 2013.79 & 3.06$\pm$0.15 & 69$\pm$5  & $<$4.58 \\
56615 & 2013.88 & 2.67$\pm$0.13 & 38$\pm$2  & $<$0.29  \\
56650 & 2013.98 & 2.53$\pm$0.13 & 47$\pm$2  & $<$0.92 \\
56716 & 2014.16 & 1.58$\pm$0.08 & 22$\pm$5  & $<$0.09  \\
56738 & 2014.22 & 1.21$\pm$0.06 & 25$\pm$7  & $<$0.33  \\
56769 & 2014.30 & 1.90$\pm$0.10 & 50$\pm$3  & $<$2.28 \\
56902 & 2014.67 & 3.03$\pm$0.15 & 39$\pm$4  & $<$0.28  \\
56960 & 2014.83 & 2.26$\pm$0.11 & 28$\pm$1  & $<$0.09  \\
56990 & 2014.91 & 3.56$\pm$0.18 & 38$\pm$3  & $<$0.17  \\
57038 & 2015.04 & 2.71$\pm$0.14 & 64$\pm$1  & $<$3.72 \\
57076 & 2015.15 & 2.22$\pm$0.12 & 22$\pm$2  & $<$0.03  \\
57107 & 2015.23 & 2.36$\pm$0.12 & 29$\pm$3  & $<$0.10  \\
57318 & 2015.81 & 2.15$\pm$0.11 & 57$\pm$2  & $<$3.37 \\
57356 & 2015.91 & 1.72$\pm$0.09 & 41$\pm$3  & $<$1.04 \\
57384 & 2015.99 & 1.27$\pm$0.06 & 31$\pm$5  & $<$0.61  \\
57429 & 2016.11 & 1.75$\pm$0.09 & 52$\pm$3  & $<$3.25 \\
57448 & 2016.16 & 1.24$\pm$0.06 & 28$\pm$2  & $<$0.30  \\
\hline
\end{tabular}
\end{table}
\end{document}